\documentclass{aa}

\usepackage{graphicx}
\usepackage{txfonts}
\usepackage{siunitx}
\usepackage{xcolor}

\usepackage[normalem]{ulem}

\usepackage{hyperref}
\hypersetup{%
  colorlinks=true,
  linkcolor=blue,
  urlcolor=blue,
  citecolor=blue
  }

\begin{document}

\newcommand{\hrieuv}{HRIEUV\xspace}
\newcommand{\hrilya}{HRILy$\alpha$\xspace}

\newcommand{\fsieuv}{FSI 174 \xspace}

\newcommand{\seqa}{$\mathrm{S_1}$\xspace}
\newcommand{\seqb}{$\mathrm{S_2}$\xspace}
\newcommand{\seqc}{$\mathrm{S_3}$\xspace}

\newcommand{\eva}{$\mathrm{E_1}$\xspace}
\newcommand{\evb}{$\mathrm{E_2}$\xspace}
\newcommand{\evc}{$\mathrm{E_3}$\xspace}
\newcommand{\evd}{$\mathrm{E_4}$\xspace}
\newcommand{\eve}{$\mathrm{E_5}$\xspace}
\newcommand{\evf}{$\mathrm{E_6}$\xspace}
\newcommand{\evg}{$\mathrm{E_7}$\xspace}
\newcommand{\evh}{$\mathrm{E_8}$\xspace}

\newcommand{\evmethod}{$\mathrm{E_{9}}$\xspace}

\newcommand{\evlbeis}{$\mathrm{E_{10}}$\xspace}
\newcommand{\evl}{$\mathrm{E_{11}}$\xspace}
\newcommand{\evm}{$\mathrm{E_{11}}$\xspace}
\newcommand{\software}{\textit}

   \title{Spectroscopic evidence of cool plasma in quiet Sun HRIEUV small scale brightenings}

   \author{A. Dolliou \inst{\ref{aff:ias}}
          \and S. Parenti \inst{\ref{aff:ias}}
        \and K. Bocchialini \inst{\ref{aff:ias}}
          }

    \institute{%
        \label{aff:ias}{Université Paris--Saclay, CNRS,  Institut d'astrophysique spatiale, 91405, Orsay, France}
    \\ \email{antoine.dolliou@universite-paris-saclay.fr}}

   \date{Received 19 April 2024; accepted 17 May 2024}

 
  \abstract
   { A large number of the small and the short-lived EUV brightenings have been detected in the quiet Sun (QS) over the past three years, by the High Resolution Imager of the Extreme Ultraviolet Imager (\hrieuv) on board Solar Orbiter. It is still uncertain whether these events reach coronal temperatures, and thus if they directly participate to coronal heating. }
   {In this work, we evaluate the maximum temperature of 11 EUV brightenings in the QS, through plasma diagnostics involving UV/EUV spectroscopy and imaging. }
   {We use three QS observations coordinated between \hrieuv, the Spectral Imaging of the Coronal Environment (SPICE/Solar Orbiter), the EUV Imaging Spectrometer (EIS/Hinode), and the Atmospheric Imaging Assembly (AIA/SDO). We detected events in \hrieuv, ranging from \SI{0.8}{} to \SI{6.2}{\mega\meter} in length. We then identified nine of them in SPICE and AIA, as well as three in EIS. We investigated their temporal evolution using their light curves, and applied temperature diagnostics, such as the LOCI Emission Measure (EM) and the differential EM (DEM). We also estimated the electron density of one event identified in EIS.}
   {These events are dominated by emission of plasma at chromospheric and TR temperatures, and they barely reach temperatures above \SI{1}{\mega\kelvin}. As such, we concluded that their contribution to coronal heating is not dominant. The estimated density of one of the event is $n_{\mathrm{e}} =$ \SI[separate-uncertainty = true]{1.8(13)e10} {\per\centi\meter\tothe{3}}.
   }
   {}

   \keywords{Sun: corona -- Sun: transition region -- Sun: UV radiation -- Instrumentation: high angular resolution -- Instrumentation: spectrographs 
               }
    \titlerunning{Spectroscopic evidence of cool plasma in quiet Sun HRIEUV small scale brightenings}
    \authorrunning{Dolliou et al.}
   \maketitle
%

\section{Introduction}

The corona is the outer layer of the solar atmosphere. It is maintained at temperatures above \SI{1}{\mega\kelvin} through processes only partially understood. For this, wave heating and magnetic reconnexions are predicted to play a major role. See, for instance, \cite{Reale2014}, \cite{VanDoorsselaere_2020} and \cite{Viall21} for reviews on these arguments.

Observations in active regions (AR) indicate that the heating is impulsive in nature \citep{Lundquist_2008,Ugarte_2019} and that most of it happens at unresolved spatial scales \citep{Hudson1991}. One of the main theories \citep{Parker1988} suggests that the corona is formed through a large number of small-scale ($\sim 10^{24}$ erg) and impulsive heating events called "nanoflares". They are the consequence of reconnexion  of magnetic fields. The definition of a nanoflare has evolved since then \citep[][]{Klimchuk_2015}. They can now refer to any impulsive and small-scale energy dissipation, regardless of the physical mechanism responsible for it, such as magnetic reconnexion or wave. Because of the unresolved size of the nanoflares, this heating has been investigated in active regions for decades through the comparison of observational signatures \citep[i.e ][]{Cargill_1997,Parenti_2017} and modellings \citep[i.e.][]{ Parenti_2006,Parenti_2008,Bingert_2011,Bradshaw_2012,Cargill_2014}. While most of the studies focused on AR, evidence of waves \citep[i.e.][]{McIntosh2011,Hahn&Savin2014} and magnetic reconnexion \citep[i.e.][]{Tripathi_2021,Upendran2021,Kahil_2022} have also been found in the quiet Sun (QS).

 Observations showed that the energy distribution of X-ray microflares to flares \citep{Crosby_1993,Hannah_2008} and EUV brightenings \citep{Berghmans_1998,Aschwanden_2000,Joulin_2016} follows a power law, with the lowest-energy events being the more frequent. As of now, all of the detected events can not, by themselves, explain coronal heating \citep{Hudson1991,Aschwanden_2002}. It has been speculated that impulsive events at even shorter and unresolved scales might provide the necessary power to heat the corona. The new generation of instruments, observing at high spatial and temporal resolutions, confirms what was predicted by the impulsive heating models: more and more short-scale events are detected in the EUV. Example of them include small (\SI{7.5}{\mega\meter}) and cool (\SI{2.5e5}{\kelvin}) loops \citep{Winebarger2013} detected with the High-Resolution Coronal Imager (Hi-C) sounding rocket flights \citep{Kobayashi2014}. Likewise in Hi-C, \cite{Regnier2014} detected \SI{680}{\kilo\meter} length and 25-second duration "EUV bright dots" at the edge of active regions. \cite{antolin_reconnection_2021} identified small (\SI{1}{} to \SI{2}{\mega\meter}) and very short lived (up to \SI{15}{\second}) "nanojets" in large coronal loops, with the Interface Region Imaging Spectrograph \citep[IRIS; ][]{DePontieu2014}, and the Atmospheric Imaging Assembly \citep[AIA; ][]{Lemen2012}, on board the Solar Dynamics Observatory \citep[SDO;][]{Pesnell2012}. \cite{Young2018} reviewed the properties of UV bursts observed by AIA and IRIS. In addition to EUV, X-ray bursts associated with smaller flares have been extensively studied \citep{Hannah_2010,Hannah_2019,Buitrago_2022}, as such emission could provide enough energy to heat the corona \citep[for a review, see ][]{Hannah_2011}. 

The Solar Orbiter mission \citep[][]{Muller,Zouganelis2020}, launched in February 2020, carries, among others, the Extreme Ultraviolet Imager \citep[EUI;][]{EUI_instrument} and the UV Spectral Imaging of the Coronal Environment \citep[SPICE;][]{Anderson_2020}.  EUI includes the Extreme Ultraviolet High Resolution Imager (\hrieuv), with a response function centred around the \ion{Fe}{X} \SI{174.5}{\angstrom} line (\SI{1}{\mega\kelvin}). One of the novelties of this mission is to approach the Sun at 0.28 AU. At this distance, the spatial resolution of \hrieuv reaches about \SI{200}{\kilo\meter}. Such resolution was confirmed in-flight by \cite{Berghmans_2023}, who measured the full width at half maximum (FWHM) of dot-like features in the corona. The average FWHM over all the features was equal to 1.5 pixels, which showed that the spatial resolution of \hrieuv is limited by the two-pixels Nyquist limit. The combination of the temperature sensitivity and the high spatial cadence makes \hrieuv and SPICE suitable instruments to study coronal heating.

On 2020 May 30, \hrieuv made its first high cadence \SI{5}{\second} observation of the QS, while Solar Orbiter was at 0.558 AU from the Sun. During the \SI{4}{\minute} sequence, \cite{Berghmans2021} automatically detected 1468 small (\SI{400}{} to \SI{4000}{\kilo\meter}) and short-lived (\SI{10}{} to \SI{200}{\second}) EUV brightenings (here called events). They are low in the atmosphere, between \SI{1000}{} and \SI{5000}{\kilo\meter} above the photosphere \citep{Zhukov2021}, and they are concentrated around the chromospheric network.

In this paper, we aim at verifying if, indeed, these events contribute to the heating of the corona. If that is the case, they should, at least, reach coronal temperatures. However, temperature diagnostics were challenging on 2020 May 30, as no spectroscopic data were available at that time. Instead, the \hrieuv field of view was within that of AIA, which had its coronal channels running at a \SI{12}{\second} cadence. \cite{Dolliou_2023} used the AIA channels to perform temperature diagnostics on the same dataset. They looked for delays between the AIA light curves, that could be signatures of plasma cooling from temperatures above \SI{1}{\mega\kelvin} \citep[see also][]{Viall&Klimchuk2016}. They concluded that the event dataset might be dominated by brightenings not reaching temperatures above \SI{1}{\mega\kelvin}. Another possibility raised by the authors, was that AIA did not temporally resolve the fast plasma cooling from coronal temperatures. In the present study, we demonstrate that these events are dominated by plasma below \SI{1}{\mega\kelvin}.

Only a few studies, up to now, have investigated these events using spectroscopy. \cite{Huang_2023} identified three EUV brightenings in \hrieuv and in SPICE on QS observations. The events had emission in the chromospheric (\ion{C}{III}, $\log{T} = 4.8$) and the transition region (\ion{O}{VI}, $\log{T} = 5.5$) lines, and two of them were detected in the \ion{Ne}{VIII} line ($\log{T} = 5.8$). \cite{Nelson_2023} also identified EUV brightenings in the QS with \hrieuv and IRIS. The events showed strong and diverse responses in the chromospheric lines. The authors concluded that these EUV brightenings could include a large variety of events with multiple physical origins. These two studies provided results consistent with events barely reaching coronal temperatures.

In this work, we apply spectroscopic diagnostics to multiple EUV brightenings detected by \hrieuv. To cover the required wide temperature range, we analyse data from SPICE and, for the first time, with brightenings detected by \hrieuv, the EUV Imaging Spectrometer \citep[EIS,][]{Culhane_2007}, on board the Hinode spacecraft \citep[][]{Kosugi_2007}. Section \ref{sec:observation} presents the QS observations with \hrieuv, SPICE, AIA, and EIS. Section \ref{sec:method} details the events detection with \hrieuv, their identification in the other instruments, and the different diagnostics applied. We show that the events are dominated by plasma at chromospheric and TR temperatures in Sect. \ref{sec:results}. We discuss the implication of these results in Sect. \ref{sec:discussion}.


\section{Observations and data reduction}
\label{sec:observation}

\renewcommand{\arraystretch}{1.5} 
\begin{table*}
\caption{ Detailed information on the datasets used in this work. $D_\mathrm{Sun}$ is the distance between Solar Orbiter and the centre of the Sun. $\theta$ is the separation angle between the Solar Orbiter view and the Earth perspective. The number of events refers to the events listed in Table \ref{table:event_properties}. $\Delta t$ is the observation cadence. When no time is indicated in the "Time" column, this means that the observation is available during the whole \hrieuv sequence.}
\label{table:instrument_sequences_S123}      
\flushleft          
\begin{tabular}{c c c c c | c c c c c }      
 \hline 
 \hline 
  \multicolumn{5}{c|}{Sequence} & \multicolumn{4}{c}{Instruments}  \\
   \hline 
 Dataset & Date & $D_\mathrm{Sun}$ [AU] & $\theta$ [$^{o}$] & \# Events & Name & Time [UTC] & $\Delta t$ [s]  & Comment \\
\hline                    
  \seqa  & 2022 March 30  & 0.487 & 0.8 & 8 & \hrieuv & 00:00\,--\,03:00 & 3 to 20  & \\
  
   & & & & & AIA &  & 6 &  &  \\
   & & & & & SPICE & 00:00\,--\,02:43 & 25 & three-step raster  \\
\hline
   \seqb & 2022 March 17 & 0.380 & 25.9 & 1 & \hrieuv & 00:18\,--\,00:48 & 3 & \\
   & & & & & AIA & & 6  & \\
   & & & & & SPICE & & 25  & three-step raster \\
   \hline 
   \seqc & 2023 April 4 &  0.326 & 27.5 & 2 & \hrieuv & 04:23\,--\,06:56 & 3 & \\
   & &  & & & AIA & 04:23\,--\,06:56 & 12 & \\
   & &  & & & EIS & 04:23\,--\,05:25 & 3720 & 60-step raster \\
   & & & & & EIS & 06:01\,--\,06:56 & 414 & 25-step raster \\ 
\hline       
\\
\end{tabular}

\end{table*}

\begin{figure*}[ht]
    \centering
    \includegraphics{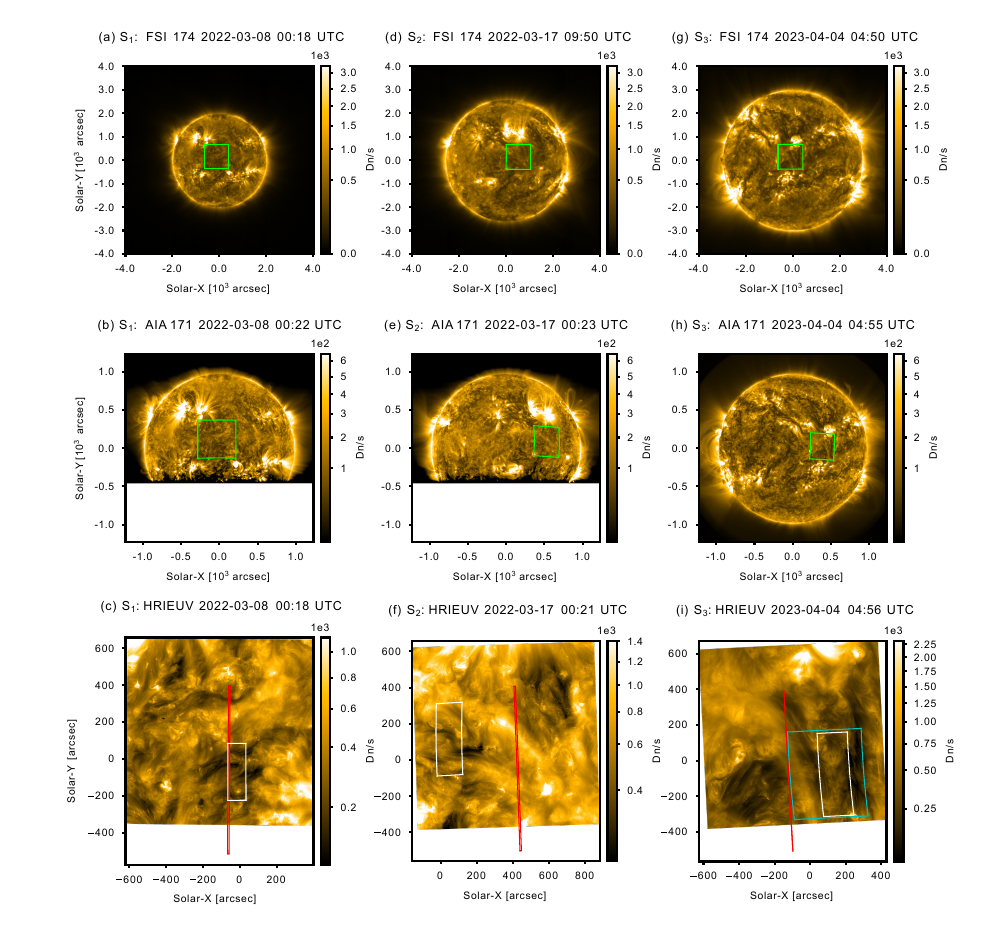}
        \caption{Quiet Sun datasets on 2022 March 8 (first column), 17 (second column) and on 2023 April 4 (third column). Upper row: field of view of \hrieuv (green) within the \fsieuv image. Middle row: field of view of \hrieuv as seen by AIA 171. The AIA images are cropped in \seqa (b) and \seqb (e), to save reading time of the detectors during the high cadence mode.  Lower row: \hrieuv image with the field of view of the SPICE (red), the EIS context (white) and the EIS 25-step  (cyan) rasters.  SPICE is in a three -step raster mode on 2022 March 8, 17, and in sit and stare mode on 2023 April 4. }
    \label{fig::fig_02_01_HRIEUV_FSI_SPICE_EIS_fov}
\end{figure*}

We use datasets from the Solar Orbiter Observing Plan (SOOP) called 'R\_BOTH\_HRES\_HCAD\_Nanoflares'. The objective of this SOOP is to capture the fast temporal variations of small and short-lived EUV brightenings, and to compare their properties among the quiet Sun (QS), active regions (AR) and coronal holes (CH). The SOOP is made of coordinated observations between Solar Orbiter instruments \citep{Auchere_2020,Zouganelis2020}. Here, we use datasets from \hrieuv and SPICE. The observations were also made in collaboration with instruments from other Earth-perspective observatories, including AIA and EIS. 

We selected three QS observations, which we call \seqa, \seqb, and \seqc. Table \ref{table:instrument_sequences_S123} provides the information for the three observational sequences. Detailed descriptions of the instruments and the data are given in the following section for \hrieuv (Sect. \ref{sec:observation:hrieuv}),  SPICE (Sect. \ref{sec:observation:spice}),  AIA (Sect. \ref{sec:observation:aia}) and EIS (Sect. \ref{sec:observation:EIS}).

 To identify and follow the evolution of small and short-lived events in all of the instruments, we need a careful co-registration. We developed tools for the co-registering of the \hrieuv and the SPICE sequences. They are regrouped within the \software{euispice\_coreg} Python package\footnote{\url{https://github.com/adolliou/euispice_coreg}, consulted on 2024 April 16}, and the details of the method are given in 
 Appendix \ref{sec:annex:alignment}. 

\subsection{EUI}
\label{sec:observation:hrieuv}

EUI is the UV imager on board Solar Orbiter. It includes a Full Sun Imager (FSI) at two passbands and two high-resolution imagers (HRI). The two filters of FSI are centred around the \ion{Fe}{X} \SI{174.53}{\angstrom} (FSI 174) and the \ion{He}{II} \SI{303.78}{\angstrom} (FSI 304) lines. HRI consists of two telescopes, namely \hrilya and \hrieuv.

 The \hrieuv response function peaks at \SI{1}{\mega\kelvin}, but includes a contribution from transition-region (TR) emission. The CCD detector contains 2048$\times$2048 pixels, with a pixel size of 0.492$^{\prime \prime}$, resulting in field of view of $1007.6^{\prime \prime} \times 1007.6^{\prime \prime}$. The pixel size on the corona can be derived from the Solar Orbiter--Sun distance (Table \ref{table:instrument_sequences_S123}). At 1.004~\(\textup{R}_\odot\) from the centre of the Sun, it is equal to \SI{172}{\kilo\meter}, \SI{134}{\kilo\meter}, and \SI{115}{\kilo\meter} during \seqa, \seqb, and \seqc respectively. In addition to \hrieuv, we also use both FSI imagers to co-register \hrieuv and SPICE. \fsieuv and FSI 304 have a pixel size of $4.44^{\prime \prime}$, and a field of view of $224.96^{\prime}\times 227.34^{\prime}$.

We use Level 2 (L2) FITS files from the EUI Data Release 6.0 \citep{euidatarelease6}. The data processing of the EUI FITS files from level 1 (L1) to L2 includes a dark-frame correction, a flat-field correction, and a normalisation by the integration time. The pointing information in the FSI files metadata is also corrected with a limb-fitting method. As this method cannot be applied to imagers that do not observe the full Sun, there are uncertainties in the pointing information of \hrieuv. Therefore, we co-register the \hrieuv sequence with FSI using a cross-correlation technique. Further details are given in the Appendix \ref{sec:annex:alignment}.

Figure \ref{fig::fig_02_01_HRIEUV_FSI_SPICE_EIS_fov} shows the field of view of \hrieuv  on the Sun as imaged by \fsieuv (top row) and AIA 171 (middle row).  An \hrieuv image (bottom row) is also displayed for each observation sequence. The fields of view cover QS areas.

\subsection{SPICE}
\label{sec:observation:spice}
\renewcommand{\arraystretch}{1.5} 
\begin{table}
\caption{SPICE lines in the three-step raster sequences and the context rasters in \seqa and \seqb. $\lambda$ is the theoretical position of the line, and $T$ is the temperature of the maximum emission.  }    
\flushleft          
\begin{tabular}{l c c} 
\hline
\hline
Ion & $\lambda$ [\SI{}{\angstrom}] & log($T$ [K]) \\
\hline
\ion{H}{I} Ly-$\beta$ & 1025.72 &  4.0 \\
\ion{C}{III} & 977.03 & 4.8 \\
\ion{S}{V} & 786.47 & 5.2 \\
\ion{O}{IV} & 787.72 & 5.2 \\
\ion{O}{VI} & 1031.93 & 5.5 \\
\ion{Ne}{VIII} & 770.42 & 5.8 \\
\ion{Mg}{IX} & 706.02 & 6.0 \\ 
\end{tabular}         
\label{table:spice_lines}  
\end{table}

SPICE is a UV spectro-imager that images the solar spectrum over two CCDs, called the short wavelength (\SI{704}{} to \SI{790}{\angstrom}) and the long wavelength  (\SI{972}{} to \SI{1049}{\angstrom}) detectors. When imaging \seqa and \seqb, SPICE was running  a three-step raster with a 25-second cadence. In addition to this, each dataset also includes a 96-step context raster. The pixel size is $4^{\prime \prime}\times 1^{\prime \prime}$. The slit chosen was $4^{\prime \prime}$ wide, and the resulting field of view for the three-step raster is $12^{\prime \prime} \times 832^{\prime \prime}$. The spatial resolution of SPICE was measured in-flight during the commissioning phase, as the FWHM of the point spread function (PSF). It was estimated to be between  6$^{\prime\prime}$ and 7$^{\prime\prime}$ \citep{Fludra_2021}, which is about six to seven times higher than the spatial resolution of \hrieuv \citep[see also][]{Plowman_2023}. The spatial resolution of SPICE must therefore be taken into consideration when identifying \hrieuv events with SPICE. Multiple small brightenings captured by \hrieuv, close to one another, may be unresolved in the SPICE images.

  Spectrally, each series contains selected line profiles listed in Table \ref{table:spice_lines}. They cover a temperature range from $\log{T} = 4.0$ to $6.0$. The three-step rasters share the same spectral windows as the context raster.

We use L2 data for SPICE, taken from the Data Release 4.0 \citep{spice_data_release_4.0}. The processing from the L1 to the L2 FITS files includes a initial pointing correction, a dark subtraction, a flat-fielding, a burn-in correction, a distortion correction and a radiometric calibration. We also apply a sigma-clipping algorithm to remove possible spikes, that can be caused by cosmic rays. Similarly to \hrieuv, there are uncertainties in the pointing information of the SPICE FITS files, which must be corrected. For each dataset, we co-align the SPICE context raster with either FSI 304 or \hrieuv, depending on the data availability. The details about the procedure are given in Appendix\,\ref{sec:annex:alignment}. As a second step, the three-step rasters are then co-registered, as they have a fixed and known position with respect to the context raster. 

The uncertainties in the intensity of the SPICE pixels include the contributions from the dark current, the read noise and the shot noise \citep[see also][]{Huang_2023}. They are computed with the SPICE data-analysis Python package \software{sospice\footnote{\url{https://github.com/solo-spice/sospice}, consulted on 2024 April 16}}. The spectral lines are fitted with Gaussian functions using the \software{SolarSoft SPICE data analysis software\footnote{\url{https://github.com/ITA-Solar/solo-spice-ql}, consulted on 2024 April 16}}. The software returns the values and the uncertainties of the fitting parameters. The fitted parameters are then used to calculate the radiance of the line. 

To constrain our results for the plasma-temperature analysis, we used the information that some lines are not detectable in the spectra. This will be used to set an upper limit to the radiance of the regions where the spectrum is averaged. When no line is detected, we assume a radiance equivalent to that of a line with an amplitude of 2$\sigma$ and a width given by the median of the line widths in the whole context raster. Here, $\sigma$ is the combination of the dark and the read noise \citep[see also][]{Parenti_2017}. In this work, the radiance is calculated over an extended area including several pixels (see section \ref{sec:method:event_detection}). In this case, the noise is defined as the root sum square of the noise in the pixels of that region, divided by the number of pixels.

The SPICE data include a sit-and-stare sequence on \seqc.  The field of view is displayed in Fig. \ref{fig::fig_02_01_HRIEUV_FSI_SPICE_EIS_fov} (c). No clear signature of events was seen in the data. We therefore consider that no events were detected by SPICE.

\subsection{AIA}
\label{sec:observation:aia}

AIA is a EUV--UV imager that includes  six EUV channels distributed within four telescopes. For \seqa and \seqb, AIA was available in a special six-second cadence mode, to follow the \hrieuv high cadence as close as possible. The objective was to capture the full thermal evolution of the EUV brightenings detected by \hrieuv. Only four channels (171, 131, 193, 304) are available during this mode. Figures \ref{fig::fig_02_01_HRIEUV_FSI_SPICE_EIS_fov} (b), (e) shows an AIA 171 image taken from the \SI{6}{\second} cadence mode, respectively for \seqa and \seqb. The bottom of the AIA images has been cropped during the special mode to allow the fast reading of the detector and to keep the high cadence. Full-Sun images were also obtained every \SI{96}{\second}. Figures \ref{fig::fig_02_01_HRIEUV_FSI_SPICE_EIS_fov} (h) shows an AIA 171 image taken during \seqc, when AIA was not running the high-cadence mode.

The level 1 AIA datasets are prepared with the \software{aiapy open project \citep{Barnes2020}\footnote{\url{https://gitlab.com/LMSAL_HUB/aia_hub/aiapy}, consulted on 2024 April 16}}. First, the pointing information is updated. Then, the images are deconvolved with the PSF estimated by the Richardson--Lucy deconvolution algorithm. Finally, the roll angle of the images is removed and the degradation in time of the CCD is corrected.

We project the AIA images on the \hrieuv pixel grid using \software{sunpy  \citep{sunpy_community2020}} and the \software{reproject} library from \software{Astropy \citep{astropy:2013, astropy:2018, astropy:2022}}.  The AIA 171 images are then co-aligned with the \hrieuv images. This step removes the average shift caused by the separation angle between Solar Orbiter and SDO. The resulting shift correction is applied to the AIA 193, 131, and 304 channels.

\subsection{EIS}
\label{sec:observation:EIS}

EIS is an EUV spectrometer imaging the solar spectrum on a short-wavelength (\SI{170}{} to \SI{210}{\angstrom}) and a long-wavelength detector (\SI{250}{} to \SI{290}{\angstrom}).  The \seqc dataset is made of a context raster and a faster 25-step raster sequence. The slit is 2$^{\prime \prime}$ wide, with a 1$^{\prime \prime}$ pixel size along the slit. At 1.004~\(\textup{R}_\odot\) from the centre of the Sun, the pixel size corresponds to about \SI{1441}{\kilo\meter} and \SI{721}{\kilo\meter} in the Solar corona. The spatial resolution of EIS is estimated to be around $3^{\prime \prime}$\footnote{\url{https://sohoftp.nascom.nasa.gov/solarsoft/hinode/eis/doc/eis_notes/08_COMA/eis_swnote_08.pdf}, consulted on 2024 April 16}. The context raster acquired the full spectrum, while the smaller ones only covered selected lines (see Table\,\ref{table:eis_line_radiance}). In this study, we use lines emitted from chromospheric (\ion{He}{II} \SI{256.32}{\angstrom}) to coronal  (\ion{Fe}{XVI} \SI{262.98}{\angstrom}) temperatures.

 The EIS data reduction and analysis are performed with the \software{eispac} Python package\footnote{\url{https://github.com/USNavalResearchLaboratory/eispac}, consulted on 2024 April 16}. The data have been previously prepared with the \textit{"eis\_prep"} routine available via \software{solarsoft} in  \software{Interactive Data Language} (IDL) . During the preparation, the CCD bias and the dark-current background are subtracted. The cosmic rays, along with the hot and the cold pixels are flagged. The wavelengths are corrected for the orbital motion, the temperature effect, and the tilt of the slit relative to the CCD. We also correct the pointing information with the co-registration algorithm developed by \cite{Pelouze_2019}. We also use the recent radiometric calibration of \cite{Del_Zanna_2023_submitted}, which corrects for the temporal degradation of the CCD. Using \software{eispac}, we perform line fitting with one or multiple Gaussian functions, to compute the line radiance.

Figures \ref{fig::fig_02_01_HRIEUV_FSI_SPICE_EIS_fov} (c), (f), and (i) display the field of view of EIS on an \hrieuv image for \seqa, \seqb, and \seqc respectively. We can see that the uncertainties between the instruments or spacecraft pointings resulted in no overlap between the fields of view of EIS and SPICE. For this reason, no event could be observed by SPICE and EIS simultaneously. We then did a separate analysis.

 While EIS datasets were available for all three observation sequences, we only use the one for \seqc, because no \hrieuv events could be seen in EIS. One possible reason is that both datasets had lower exposure time (10 and \SI{30}{\second}) compared with that of \seqc (15 and \SI{60}{\second}), for the 25-step and the context rasters respectively.


\section{Method}
\label{sec:method}
\begin{figure*}
    \centering
    \includegraphics[width=\textwidth]{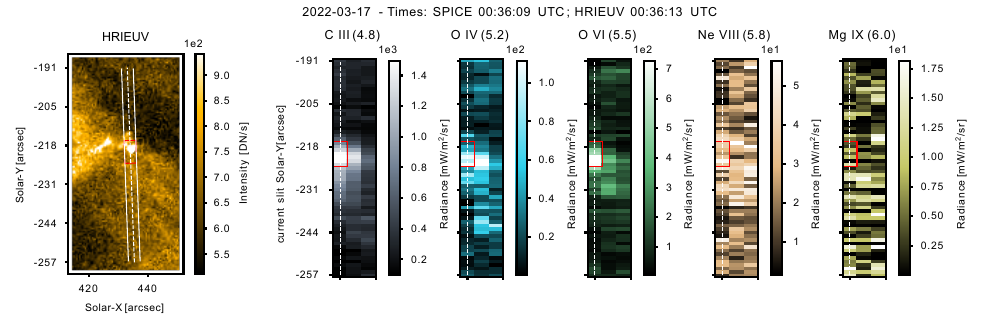}
    \caption{Event detected on 2022 March 17 (\seqa), so-called \evmethod. The \hrieuv (left panel) and the SPICE images are centred around \evmethod. The \hrieuv image is taken at the acquisition time of the left-hand SPICE slit. The position of the slit is displayed as a dashed line on the \hrieuv and the SPICE images. The two white lines in \hrieuv delimit the field of view of the SPICE slit. The red rectangle is the event region, defined in Sect. \ref{sec:method:identification_SPICE}. The  temperature ($\log{T}$) of the maximum emissivity of each SPICE line is indicated within parentheses above  each image. The latitude on the $y$-axis of the SPICE images refers to the position of the slit marked with a white dashed line.
    }
\label{fig:method:bright_dot_170322_all_event_spice_imshow_peak2}   
\end{figure*}

\begin{figure*}
    \centering
    \includegraphics[width=\textwidth]{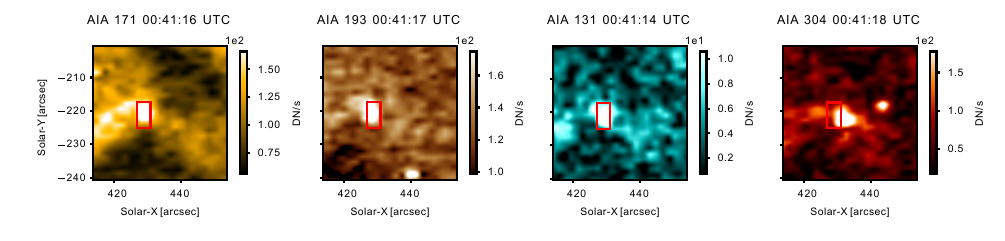}
    \caption{Images of four AIA channels centred around the location of \evmethod. The images are the closest in time to the \hrieuv image in Fig. \ref{fig:method:bright_dot_170322_all_event_spice_imshow_peak2}, taking into account the delay caused by the light travel time. The red rectangle is the event region.}
\label{fig:method:bright_dot_170322_all_event_aia_imshow_peak2}
\end{figure*}

\begin{figure*}
\centering
    \includegraphics[width=\textwidth]{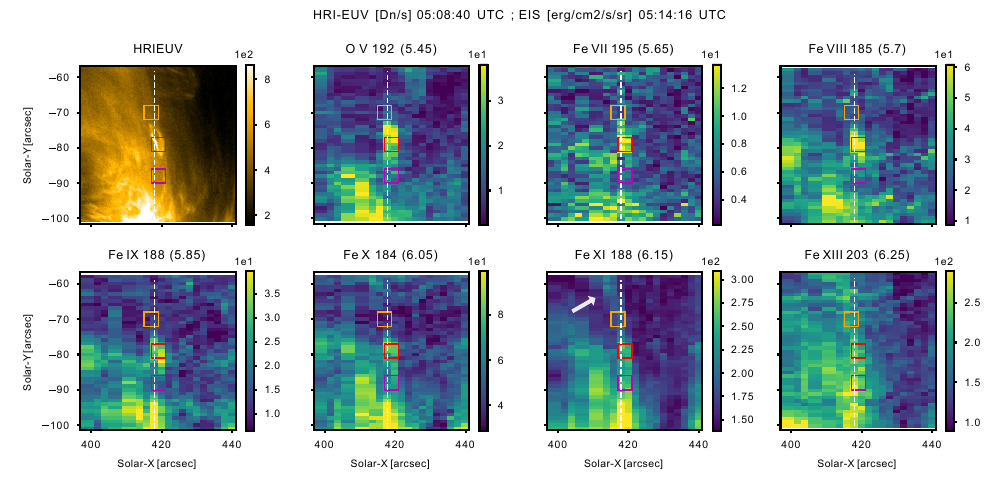}
    \caption{ \hrieuv (top-left panel) and EIS images centred around the event \evlbeis, on 2023 April 4 (\seqc). The \hrieuv image is taken at the  time closest to that of the EIS slit position indicated by dashed white lines. The event region (red), along with the background regions 1 (orange) and 2 (purple) are displayed as rectangles. These regions are defined in Sect. \ref{sec:method:identification_EIS}.  The temperature $\log{T}$ of the maximum emissivity of each line is indicated inside parenthesis on top of the image. The white arrow in the \ion{Fe}{XI} image indicates the large coronal loop standing above \evlbeis.}
    \label{fig:method:eis_large_brightening_imshow}
\end{figure*}

 The events were detected in \hrieuv data using the automated wavelet detection code described by \cite{Berghmans2021}. We then selected a sub-population of the events in the overlapping region observed by the different instruments. We did not include events that were located only partially  within the field of view, or events that were not clearly detected in either EIS or SPICE. Many of the smaller events detected in \hrieuv could not be identified in EIS, while most of them were identified in SPICE. As a matter of fact, we could only report two events that showed no signature in SPICE. These two events are displayed in Figs. \ref{fig:annex:method:event_undetected_spice1} and \ref{fig:annex:method:event_undetected_spice2}. They are discussed in Sect. \ref{sec:discussion}, and they are not included in our dataset.  Table \ref{table:event_properties} lists the final dataset which consists of eight events in \seqa, one in \seqb, and two in \seqc. 

Table \ref{table:event_properties} presents their given name, an estimation of their lifetime and of their apparent length. The lifetime is set by the period when the event is visible above the background emission. We measure the apparent length $l$ in the \hrieuv image at the time of the intensity peak. \evm was a special case, as it had a resolved loop-like shape. In that case, we measured the length along the loop $L \approx \frac{1}{2}\pi l$. Both $L$ and $l$ are lower limits, as the loop may be tilted with respect to the line of sight, and it might only be partially visible in the \hrieuv passband. 

  During their lifetime, some of the observed events show multiple peaks in \hrieuv intensity. These multiple peaks may share the same physical origin, or they can be unrelated. As the apparent shape of the events can change between the peaks, we analyze the peaks separately unless specifically mentioned. 
  
  The event detection with \hrieuv and their identification in SPICE and EIS are discussed in Sect. \ref{sec:method:event_detection}. The diagnostics including the light curves, the emission measure (EM) LOCI, the differential EM, and the line ratios are described in Sect. \ref{sec:method:plasma_diagnostic}.

\subsection{Event detection}
\label{sec:method:event_detection}

\begin{table}
\caption{Events identified in \hrieuv and in either SPICE (\eva to \evmethod) or EIS (\evlbeis and \evm) data. From left to right, the columns present the central time in UTC, the name, the lifetime $\tau$, the apparent length $l$, and a short description of their apparent shape. The "on a structure" description indicates that the event is located on a larger structure.}           
\label{table:event_properties} 
\flushleft          
\begin{tabular}{c c c c c} 
\hline
\hline
 Central time  & Name & $\tau$ [s] & $l$ [Mm] & Description \\
\hline  
\multicolumn{5}{c}{2022 March 8 (\seqa)} \\
\hline
  00:39:03 & \eva & 70 & 0.9 &  dot-like \\ 
   00:40:13 & \evb & 60 & 1.0 &  on a structure \\ 
   01:23:58 & \evc & 420 & 1.2 & on a structure \\
   01:31:03 & \evd & 65 & 1.0 &  loop-like \\
  00:52:23 & \eve & 75 & 2.2 &  loop-like \\ 
    02:03:01 & \evf & 275 & 1.5 & complex  \\ 
  01:37:18 & \evg & 216 & 2.6 &  complex \\
 02:25:38 & \evh & 856 & 2.5 &  loop-like \\
\hline  
\multicolumn{5}{c}{2022 March 17 (\seqb)} \\
\hline
  00:36:09 & \evmethod & 440 & 0.8 &  on a structure \\ 
\hline  
\multicolumn{5}{c}{2023 April 4 (\seqc)} \\
\hline
  05:08:27 & \evlbeis & 1140 & 6.1 &  complex \\ 
   06:30:18 & \evm & 270 & 3.2 &  loop-like/jet \\ %
\end{tabular}

\end{table}

\subsubsection{Identification of the events within SPICE and AIA}
\label{sec:method:identification_SPICE}

In \seqa and \seqb images, we identified nine \hrieuv events both in SPICE and AIA. In this section, we provide the information on how we identified these events, and how we build the light curves that will be used for analysis. The event called \evmethod, detected in \seqb, will be used to illustrate the analysis steps. Section \ref{sec:results:spice} discusses the results for three events including \evmethod, while Appendix \ref{annex:080322_events} presents the results for the rest of the events. Examples of events seen by \hrieuv, SPICE and AIA are also available as movies in the multimedia materials.

Figure \ref{fig:method:bright_dot_170322_all_event_spice_imshow_peak2} shows the \hrieuv and the SPICE images centred around \evmethod, at the time of an \hrieuv intensity peak (00:36:19 UTC). At that time, \evmethod appears in \hrieuv as a single bright dot, as part of a larger structure. Another peak in \hrieuv  is observed at 00:33:19 UTC. At this time, \evmethod appears as two distinct bright dots within the same structure. These two dots are unresolved by SPICE (see Fig.\,\ref{fig:annex:170322_bright_dot_02_peak1_00146}).

In Fig. \ref{fig:method:bright_dot_170322_all_event_spice_imshow_peak2}, the event shows strong emission in \ion{C}{III}, \ion{O}{IV}, and \ion{O}{VI}, but none in \ion{Ne}{VIII} nor in \ion{Mg}{IX}. In these wavebands, the spectrum is often dominated by noise, and no line can be detected. This is particularly true for \ion{Mg}{IX}, as it is a weak line in the QS. This property is not restricted to \evmethod; no emission in \ion{Mg}{IX} could be associated with any of events identified in SPICE. As a result, we only use this line in Sect. \ref{sec:results:spice:loci} to provide an upper limit on the event EM around $\log{T} = 6.0$. The full movie of Fig. \ref{fig:method:bright_dot_170322_all_event_spice_imshow_peak2} is available in the multimedia materials.

We used SPICE to select the area that includes the entire event (see Fig. \ref{fig:method:bright_dot_170322_all_event_spice_imshow_peak2}). Because of the different resolution of the two instruments, the area must be small enough not to include the nearby brightenings or structures seen in \hrieuv. However, the region must also be large enough to include the event during its whole lifetime. For some events (\eva, \evb, \eve, and \evh), the \hrieuv and the SPICE intensity peaks in \ion{O}{VI} are spatially shifted, despite them being co-temporal. The shifts are less than the SPICE pixel size of \SI{4}{\arcsec}$\times$\SI{1}{\arcsec}, which is within the uncertainties of the co-alignment method. We apply a spatial correction to the event's mask in \hrieuv up to the the SPICE pixel size. 

In addition to \hrieuv and SPICE, we observe the event in four AIA channels (171, 131, 193 and 304). Figure \ref{fig:method:bright_dot_170322_all_event_aia_imshow_peak2} shows the \evmethod event region   on the AIA images reprojected onto the \hrieuv pixel grid. The AIA images are those closest in time to the \hrieuv image of Fig. \ref{fig:method:bright_dot_170322_all_event_spice_imshow_peak2} (left). The difference in time between \hrieuv and AIA corresponds to the difference in light time travel between Solar Orbiter and SDO. In \seqb, the large separation angle between Solar Orbiter and the Earth perspective (Table \ref{table:instrument_sequences_S123}) produces a spatial shift on the location of \evmethod. To correct this shift, we assume that \hrieuv and AIA 171 are sensitive to similar plasma temperatures. A spatial correction is applied to the event regions in AIA. The same values are used for all of the AIA channels. The brightenings show a different morphology and position in AIA 304 compared with the other channels. This was to be expected, as the plasma emitting at chromospheric and TR temperatures might not be at the same height. As a result, they can appear at different positions on an imager (Fig.  \ref{fig:method:bright_dot_170322_all_event_aia_imshow_peak2}). The full movie of Fig. \ref{fig:method:bright_dot_170322_all_event_aia_imshow_peak2} is available in the multimedia materials.

\subsubsection{Identification of the events with EIS}
\label{sec:method:identification_EIS}

 In \seqc, we detected two events within \hrieuv images and identified them in EIS data. Figure \ref{fig:method:eis_large_brightening_imshow} shows the \hrieuv and the EIS images centred around \evlbeis. This event was captured by the context raster with a 60-second exposure time. Multiple EUV brightenings observed in \hrieuv were located on \evlbeis. Two of them were detected less than \SI{60}{\second} prior the time the EIS slit passes through the structure. \evlbeis is clearly detected in \ion{O}{V}, and in the \ion{Fe}{} lines with temperature sensitivity up to that of \ion{Fe}{XI} ($\log{T} = 6.15$). In addition to \evlbeis, a large coronal loop is also visible in the \hrieuv image and in the EIS raster in the lines emitting at $\log{T} = 6.15$ and above (see \ion{Fe}{XI} in Fig.\,\ref{fig:method:eis_large_brightening_imshow}). As the event is mostly visible in the cooler lines, and the coronal loop in the hotter lines, we infer that the loop lies above \evlbeis. Because of this, we need to isolate the emission of the event from that of the background and the foreground along the line of sight, including the loop. For this reason, we define two "background" regions (1 and 2) shown in Fig. \ref{fig:method:eis_large_brightening_imshow}. Their location is designed to be representative of the background and the foreground emission near \evlbeis, including that of the large coronal loop. 

We also identified another event, \evm,  within the 25-step raster sequence. \evm (Fig.\,\ref{fig:annex:small_loop_3_imshow}) appears as  a loop-like structure. It is clearly detected in the EIS \ion{He}{II} line, but not in any of the \ion{Fe}{} lines. We conclude that \evm does not have enough plasma emission at $\log{T} \geq 5.7$ to be detected in the $\ion{Fe}{}$ lines above the noise level, in a 15-second exposure.

\subsection{Plasma diagnostic}
\label{sec:method:plasma_diagnostic}

For all the events and their background regions, we average the intensity within the selected boxes. The lines are fitted over the averaged spectrum, to obtain the radiance. When no line is detected within the box, the radiance is being replaced by an upper limit defined in  Sect. \ref{sec:observation:spice}.  As an example, Fig.\,\ref{fig:annex:time_peak_1} displays the fitting with Gaussian functions of the SPICE spectra spatially averaged over the \evmethod event region (Fig.\,\ref{fig:method:bright_dot_170322_all_event_spice_imshow_peak2}). 

\subsubsection{Temperature diagnostics}
\label{sec:method:emloci_dem}
We aim to measure the temperature of the events, taking into account their possible multi-thermal nature. 
In the following, we assume an optically thin solar atmosphere, an electron density $n_{\mathrm{e}}$ and an hydrogen density $n_{\mathrm{H}}$. The intensity of a spectral line $I$ is given by
\begin{equation}
\label{eq:I_s}
    I = \int_L G(n_{\mathrm{e}}, T)~n_{\mathrm{e}}(s) n_{\mathrm{H}}(s)~\mathrm{d}s
\end{equation}
The emission is integrated over the line of sight $L$. The contribution function $G(n_{\mathrm{e}}, T)$ contains the relevant atomic physics for the line formation. When needed, we use the CHIANTI atomic database \citep{Dere_Chianti}, version 10.1 \citep{Del_zanna_chianti_2021,Dere_2023}, to compute $G(n_{\mathrm{e}}, T)$. Under the ionization equilibrium provided by CHIANTI, we assume the coronal abundance given by \cite{Asplund_2021}, and a typical coronal electron density of $10^{9}$\,cm$^{-3}$. The contribution functions of the SPICE and the EIS lines used in this work are displayed in Fig.\,\ref{fig:annex:goffnt}. We see that they cover a wide range of temperature. This is extremely useful to constrain the temperature diagnostics and reach the goal of this work: establishing whether the HRIEUV are coronal features.  As will be further discussed, SPICE is mostly sensitive to chromospheric and transition region lines, while the EIS spectrum contains mostly transition region and coronal lines. 

One can rewrite Eq. \ref{eq:I_s} using the ideal gas approximation. Given that in the TR and the corona the pressure gradient is much lower than the temperature and density gradients, we can assume that the density  depends only on the temperature. In that case, the DEM can be defined with a change of variable in Eq. \ref{eq:I_s}. The DEM represents the distribution, along the line of sight, of the plasma  with  temperature.
\begin{equation}
\label{eq:dem}
    \mathrm{DEM}(T) = n_{\mathrm{e}}(T) n_{\mathrm{H}}(T)\frac{\mathrm{d}s}{\mathrm{d}T}
\end{equation}
Equation \ref{eq:I_s} can be rewritten into the following.
\begin{equation}
\label{eq:I_T}
    I = \int_T G(T)~\mathrm{DEM}(T)~\mathrm{d}T
\end{equation}

The DEM can be estimated by inverting the set of Eqs. \ref{eq:I_T} from measured lines radiance. 
In our work, \evlbeis is detected in EIS lines over a wide range of plasma temperatures (Fig.\,\ref{fig:method:eis_large_brightening_imshow}), which is a suitable condition for the DEM inversion. We 
then estimate the DEM with the inversion method developed by \cite{Hannah&Kontar2012}. 

Table \ref{table:eis_line_radiance} reports the intensity values of the lines measured in the event and the background regions. We used all of the line to estimate the DEM for the three regions, except for the density-sensitive \ion{Fe}{XIII} lines. The reasons behind this removal are further discussed in Sect. \ref{sec:results:eis:dem}. The DEM is computed from $\log{T} = 5.0$ to $6.6$. For each line, we compute the ratio of the synthetic radiance obtained with the DEM, over the measured radiance. They are displayed in Table \ref{table:eis_line_radiance}, and provide an estimation of the inversion uncertainty. They are all within the 30\% uncertainties of the atomic physics \citep{Guennou_2013}, meaning that the DEM is consistent with observations.  The results of this analysis are given in Sect. \ref{sec:results:eis:dem}.

The events detected within \seqa and \seqb (\eva to \evmethod), are identified in SPICE. Contrary to EIS, the low number of SPICE lines prevents us from estimating the DEM. Instead, we use the Emission Measure (EM) LOCI method \citep[for further details]{Del_Zanna_Mason_2018}, which provides an upper limit on the DEM. The EM is defined as the integration of the DEM over the whole temperature range.
\begin{equation}
    \label{eq:em}
    \mathrm{EM} = \int_T \mathrm{DEM}(T)~\mathrm{d}T
\end{equation}
The EM LOCI method provides the EM necessary to produce a given line intensity, assuming an isothermal plasma. For a given spectral-line intensity, the LOCI curve ($\mathrm{EM_L}$) is defined as the following:
 \begin{equation}
 \label{eq:EM_LOCI}
     \mathrm{EM_L}(T) = \frac{I}{G(T)}
 \end{equation}

We compute the LOCI curves using the \ion{C}{III}, \ion{O}{IV}, \ion{O}{VI}, and \ion{Ne}{VIII} lines. \ion{Mg}{IX} is not included, as the line is never detected above noise levels. The event's EM must be isolated from the contribution of the foreground and the background emission along the line of sight. Differently from the EIS data, SPICE data contain the temporal information on the lines intensity. We use them to estimate this background emission by averaging the light curves over two time intervals of two minutes. The intervals are called B$_1$ and B$_2$, and they are set before and after the event, respectively (Fig.\,\ref{fig:results:all_3_events_light_curves}).  The results of the EM LOCI method are given in Sect. \ref{sec:results:spice:loci}, and in Appendix \ref{annex:080322_events}.

\subsubsection{Electron density diagnostics}
\label{sec:method:line_ratio}

In addition to the temperature, we estimate the density of \evlbeis  using the ratio of line intensities. In the EIS dataset, the ratio of \ion{Fe}{IX} 188.49/197.04 \citep[][b]{Young_2009_2b} and \ion{Fe}{XIII} (203.80 + 203.83)/202.04 \citep{Young_2009} are density-sensitive. The dependence on density of the \ion{Fe}{XIII} lines between $\log{n_{\mathrm{e}}} = 8$ to $10$ can be explained by the existence of a metastable level (3s$^2$3p$^2$~$^3$P$_2$) above the ground level (3s$^2$3p$^2$~$^1$P$_2$)  \citep[for more details, see][]{DelZanna&Mason2018}. In the \ion{Fe}{IX} line ratio, only the \SI{197.86}{\angstrom} is density sensitive. We could not perform similar diagnostics with the SPICE lines, as they did not provide density sensitive line ratio.  The analysis is performed with the \software{CHIANTI} database and software available via \software{SolarSoft}. The results of this analysis are given in Sect. \ref{sec:results:eis:density}.


\section{Results}
\label{sec:results}
\begin{figure}
    \centering
    \includegraphics{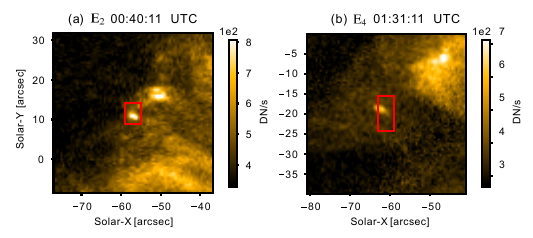}
    \caption{\hrieuv images centred around \evb (a) at the time of the second \hrieuv peak, and \evd (b) at the time of the single \hrieuv peak. The red rectangles are the event regions (see Sect. \ref{sec:method:identification_SPICE}). }
    \label{fig:results:small_dot_03_05_imshow}
\end{figure}

\begin{figure*}
    \centering
            \includegraphics[width=\textwidth]{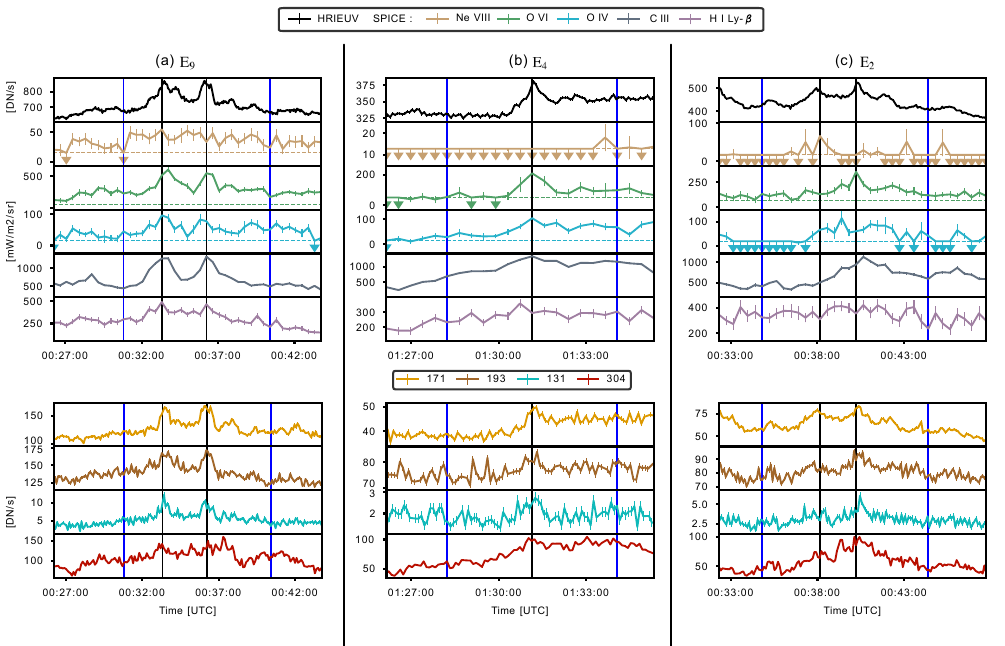}
    \caption{\hrieuv, SPICE, and AIA light curves for \evmethod (a), \evd (b), and \evb (c). The vertical black lines indicate the SPICE times closest to the \hrieuv peaks, when the EMs are estimated in Fig. \ref{fig:results:all_3_events_loci}. The two background time intervals of two minutes, defined in Sect. \ref{sec:method:emloci_dem}, are represented by two vertical blue lines in each panel. The left blue line indicates the end of B$_1$ (before the event), and the right blue line indicates the start of B$_2$ (after the event). The upper limits of the SPICE radiance are shown as horizontal dotted lines. When the line is not detected above noise levels, the radiance is being replaced by its upper limit, and the point is marked by a downward pointing arrow.
    }
    \label{fig:results:all_3_events_light_curves}
\end{figure*}

\begin{figure*}
    \includegraphics[width=\textwidth]{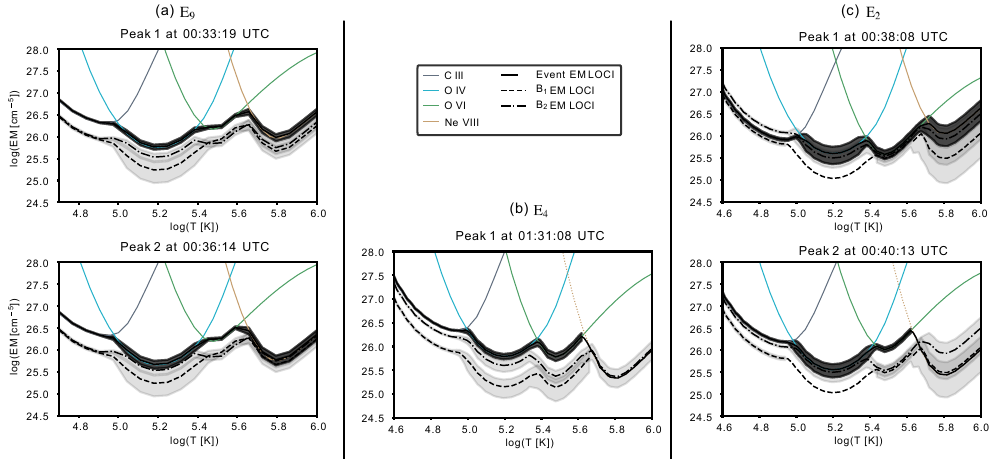}
    \caption{EM estimated with the LOCI method of \evmethod (a), \evb (b) and \evd (c) at the time of their \hrieuv peaks, using four SPICE lines. The EMs of the event are shown for each SPICE line with colored curves. Dotted curves indicate that the radiance is an upper limit. The black full curves display the EM for the event. The dotted and the dashed black lines are the EMs for the B$_1$ and B$_2$ background time intervals. B$_1$ also includes upper limits in \ion{O}{IV} for \evb and in \ion{Ne}{VIII} for \evd.  The black (event) and grey (B$_1$ and B$_2$) regions delimit the 2$\sigma$ uncertainty of the EMs.}
    \label{fig:results:all_3_events_loci}
\end{figure*}

\begin{figure*}
    \centering
    \includegraphics{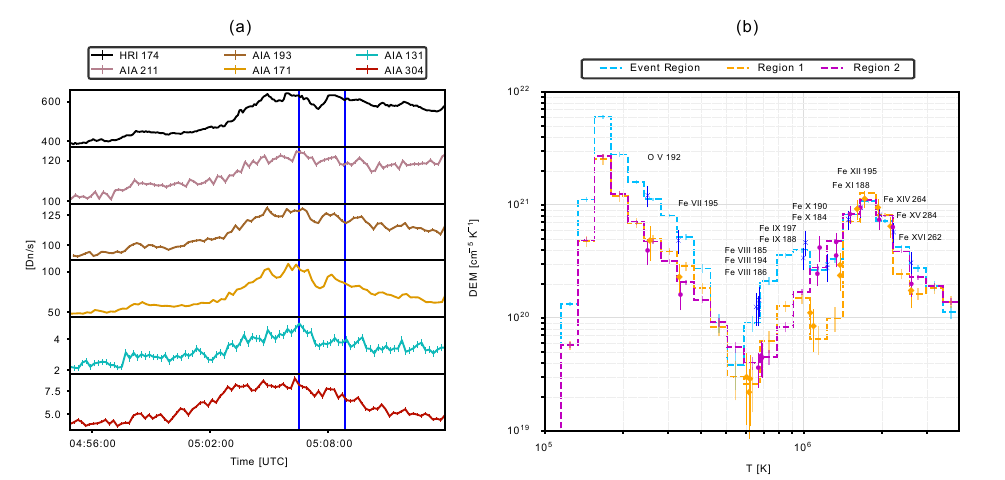}
    \caption{(a) Light curves of \hrieuv and five EUV channels of AIA with intensities averaged over the \evlbeis event region. The time interval when the EIS slit passes over the event is delimited by two blue vertical lines. (b) DEM of the event \evlbeis, along with the background regions 1 and 2. The ratios of the reconstructed radiance over the observed one is plotted on the effective temperature of each line. }

    \label{fig:results:dem_eis_coronal_no_o5}
\end{figure*}

\begin{figure}
    \centering
    \includegraphics[width=0.4\textwidth]{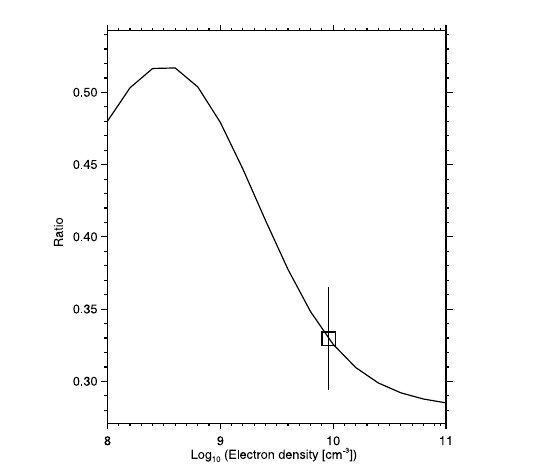}
    \caption{Line ratio \ion{Fe}{IX} 188.49/197.04 at $\log{T} = 6.0$, computed with the radiance and the uncertainty in the \evlbeis event region. The ratio provides an estimation of the electron density equal to $\log{n_{\mathrm{e}}} = 10.09 \pm 0.41$. This figure has been obtained with \software{dens\_plotter} in the \software{CHIANTI} software.}
    \label{fig:results:ratio_fe_9_event}
\end{figure}

\subsection{Transition region events detected in SPICE and AIA data} 
\label{sec:results:spice}

In the following, we analyze the temporal evolution (Sect. \ref{sec:results:spice:lc}) and obtain temperature diagnostics (Sect. \ref{sec:results:spice:loci}) of the \seqa and \seqb events. We present the results for the events \evb, \evd, and \evmethod, as they are representative of the different light curve behaviours that we encountered.  \evmethod  was already introduced in Sect. \ref{sec:method:identification_SPICE}, while \evb and \evd  are displayed  in Fig. \ref{fig:results:small_dot_03_05_imshow}  with \hrieuv, at the time of their intensity peak.  The results for the rest of the events are displayed in Appendix \ref{annex:080322_events}.

\subsubsection{Light curves}
\label{sec:results:spice:lc}

 Figure \ref{fig:results:all_3_events_light_curves} shows the \hrieuv, the SPICE, and the AIA light curves averaged over the \evb, \evd, and \evmethod regions. The events \evmethod and \evd (Fig. \ref{fig:results:all_3_events_light_curves} (a) and (b)) are visible in all  light curves, apart from that of \ion{Ne}{VIII}. Both events are also poorly detected in \ion{H}{I} Ly-$\beta$. All the intensity peaks in SPICE are co-temporal to the peaks in \hrieuv, except for a 25-second negative delay for the \ion{H}{I Ly-$\beta$} peak in \evd. Likewise, no significant delay is observed between the \hrieuv and the AIA light curves. We notice that the the \ion{O}{IV} and the \ion{O}{VI} light curves behave very similarly to the \hrieuv one.

In the event \evd (Fig. \ref{fig:results:all_3_events_light_curves} (b)), the background intensity in \hrieuv and AIA 171 increases after the event. We also observe an increase in the background intensity of the chromospheric (\ion{C}{III}, AIA 304) and TR lines (\ion{O}{IV} and \ion{O}{VI}). The increase in the \ion{C}{III} light curve is the most significant, as its value almost doubles compared with the value before the event. Due to the different temperatures of formation of these lines (Fig.\,\ref{fig:annex:goffnt}), we suggest that the common background increase is due to an increase in density at these temperatures. 

For the event \evb (Fig. \ref{fig:results:all_3_events_light_curves} (c)), the \hrieuv light curve has two peaks. The first \hrieuv peak (00:38:08 UTC) is co-temporal with the peak in the \ion{Ne}{VIII} light curve. At that time, no emission in \ion{O}{VI} is detected. The second \hrieuv peak (00:40:13 UTC) is co-temporal with the peak in \ion{O}{VI}, and it is \SI{25}{\second} earlier than the peak in the \ion{C}{III} light curve. These are indications that emission at coronal temperatures dominates at the first peak, while TR emission dominates for the second one. At 00:40:13 UTC, the cooling plasma passes through the emissivity peaks of \ion{O}{VI} followed by that of \ion{C}{III}. This example shows that the interpretation in terms of temperature of \hrieuv is not straightforward, as it images both warm and cool plasma. The interpretation of the peak in \ion{O}{IV} is uncertain, but it may be the signature of cooling plasma from the coronal temperature imaged by the first peak of \hrieuv. 

The results are confirmed by the AIA light curves of \evb (bottom rows of Fig. \ref{fig:results:all_3_events_light_curves} (c)). The two peaks have a similar amplitude in \hrieuv and in AIA 171. The amplitudes of the other AIA channels are lower at the first \hrieuv peak (00:38:08 UTC) compared with second one (00:40:13 UTC). At the first peak, the plasma probably reaches a temperature close to the sensitivity peaks of \hrieuv and AIA 171 (0.9\,--\,\SI{1}{\mega\kelvin}), but below that of AIA 193 (\SI{1.5}{\mega\kelvin}). At the second \hrieuv peak, the light-curve behavior of the AIA channels is probably caused by the TR (AIA 171, 193, 131) and chromospheric (AIA 304) contributions to the response function of the channels.

The light curves of \evb (Fig. \ref{fig:results:all_3_events_light_curves} (c)) show that, when the \hrieuv peak is caused by a plasma at coronal temperatures, we would expect the \ion{Ne}{VIII} emission to be detected in SPICE.
 In fact, under corona temperature assumption, higher \hrieuv intensity would mean higher density of the plasma at this temperature. 
The events  \eva, \evc, and \evd  all have  peaks higher than \evb. If they were caused by a coronal contribution (same temperature than \evb) they would be visible in the \ion{Ne}{viii}, and even possibly in the \ion{Mg}{ix} lines, as the EM is probably higher. We anticipate that this is not the case, as shown in Fig. \ref{fig:results:all_3_events_loci}.
This is an important result, as part of the ambiguity of the EUV brightening temperature originates from the broad response of the \hrieuv and the AIA imagers. In \evb, we have determined whether the coronal or the TR part of the response function contributed the most to the \hrieuv intensity peak. This is the evidence that the \hrieuv and AIA intensity peaks in \evmethod and \evd mostly originate from a plasma emitting at TR temperatures (Fig. \ref{fig:results:all_3_events_light_curves} (a) and (b)).

\subsubsection{EM results}
\label{sec:results:spice:loci}

Figure \ref{fig:results:all_3_events_loci} shows the results of the EM LOCI method applied to the SPICE data at the time of the \hrieuv peaks. For \evmethod and \evd, in the temperature range up to $\log{T} = 5.6$, the EM of the event is higher than that of the backgrounds (panels a, b). In particular, for the first \hrieuv peak of \evmethod, the EM of the event is higher in the whole temperature range (panel a, top). 

The EM of \evb displays different behaviors (Figure \ref{fig:results:all_3_events_loci} (c)). In the first peak (top panel), the EM of the event is similar to the one of the background. Only at temperatures above $\log{T} > 5.7$ does the EM increase a small amount.
We suggest that the \hrieuv intensity peak might be caused by the coronal lines within the band. The absence of \ion{Mg}{IX} emission above the noise level in the SPICE data can be due to an insufficient increase in the EM. We recall that this line is generally weak or absent in the QS.  At the second peak (bottom panel) of the HRIEUV light curve, the event's LOCI stands above the background EM up to $\log{T} = 5.6$. Other examples are given in Figs. \ref{fig:annex080322:E1_E3_E5_loci_article_1.00E+09} and \ref{fig:annex080322:E6_E7_E8_loci_article_1.00E+09} in Appendix \ref{annex:080322_events}.

We conclude that, in most cases, the event EM is dominated by plasma at chromopheric and TR temperatures during the \hrieuv intensity peaks. At the emissivity peak of \ion{O}{VI} ($\log{T} = 5.5$), the background-subtracted EM reaches values between \SI[separate-uncertainty = true]{3.3(13)e25}{\per\centi\meter\tothe{5}} (\evd) and \SI[separate-uncertainty = true]{8.8(21)e25}{\per\centi\meter\tothe{5}} (second peak of \evmethod). In one case (Fig.\,\ref{fig:results:all_3_events_loci} (c), top panel), we suspect the event to be dominated by coronal temperatures. Above $\log{T} = 5.6$, the event and the background EMs are close within $2\sigma$ uncertainty range, with one exception (Fig.\,\ref{fig:results:all_3_events_loci} (a), top panel). The SPICE spectrum does not include strong coronal lines. We then further investigate the temperature behavior of these brightening by analyzing EIS data, which are rich in coronal lines.  This is presented in the next section.

The absence of detectable \ion{Mg}{IX} emission in the spectra of the events places constraints on the amount of EM (and thus density) at these temperatures. We can provide an upper limit by using the upper limit of the radiance defined in Sect. \ref{sec:observation:spice} and using Eq. \ref{eq:EM_LOCI}. Taking \evmethod as an example (Fig.\,\ref{fig:results:all_3_events_light_curves} (a)), we estimate the EM to be below \SI{1.0e26}{\per\centi\meter\tothe{5}} at $\log{T} = 6.0$. Using this value, we estimate the density as $n = \sqrt{\frac{\mathrm{EM}}{h f}}$, where $h$ is the event thickness along the line of sight, and $f$ the filling factor of \evmethod over the event region \citep{Joulin_2016}. To estimate  $h$, the event is assumed to be located on a loop with a circular cross-section. $h$ is then equal to the width of the apparent loop. This width is measured by fitting a Gaussian function to the \hrieuv intensity on a perpendicular cut to the loop axis, at the position of the \hrieuv intensity peak. The width is defined as the Full Width at Half Maximum of the Gaussian function. For \evmethod, we measure $h=$ \SI{0.56}{\mega\meter}. We also estimated the filling factor $f$ using \hrieuv. To do so, we compute the ratio of the so-called "event" pixels (pixels where the event is detected) over the "QS" pixels (where the event is not detected within the selected event region). The event pixels are selected using a $3\sigma$ threshold above the average intensity $\mu$ of the QS . Both the spatial average $\mu$ and standard deviation $\sigma$ of the QS are measured within the event region, at a time prior to the event (00:30:49 UTC). At the time of the second \hrieuv intensity peak (00:36:14 UTC), we obtain a filling factor equal to $f=0.35$. Using the above expression, we derive an upper limit of the density of $n_{\mathrm{e}} =$ \SI{2.3e9}{\per\centi\meter\tothe{3}}

\subsection{Event detected with EIS}

In the following, we apply temperature and density diagnostics to \evlbeis, which is observed with EIS.  The event was captured at a single time by the 60-step context raster, preventing us from constructing light curves. Nevertheless, we could perform a density diagnostics and the differential emission measure analysis, providing more precise information on the temperature distribution of the plasma. 
The event and the background regions of \evlbeis are displayed in Fig. \ref{fig:method:eis_large_brightening_imshow}.

Figure \ref{fig:results:dem_eis_coronal_no_o5} (a) shows the light curves of \hrieuv and five EUV channels of AIA, obtained by averaging the intensity over the event region of \evlbeis. The vertical lines mark the interval of acquisition of EIS. As for the previous events (\evb, \evd, and \evmethod), there is no apparent delay between the peaks in the \hrieuv and the AIA light curves. We also see that EIS starts observing the event \SI{1}{\min} after the \hrieuv peak, meaning that the following temperature and density diagnostics are obtained close to the event peak. This event was long (\SI{19}{\min}) in \hrieuv, with multiple brightenings. Here we only concentrate on the period recorded in coordination with EIS.

\subsubsection{DEM results}
\label{sec:results:eis:dem}

We compute the DEM in the event and the background regions to infer the temperature distribution of the event. We recall that this event is located below a large scale loop, visible in the hot lines of EIS (see Fig. \ref{fig:method:eis_large_brightening_imshow}). We will take this into consideration when interpreting our results. To compute the DEM, we included all the lines listed in Table \ref{table:eis_line_radiance}, except for the density-sensitive \ion{Fe}{XIII} lines. As \evlbeis is undetected in the \ion{Fe}{XIII} lines  (Fig. \ref{fig:method:eis_large_brightening_imshow}), removing them does not impact our results.

 Figure \ref{fig:results:dem_eis_coronal_no_o5} (b) shows the DEM of the \evlbeis event  (blue line) and of the two background regions (magenta and yellow lines). For these inversions, we mostly use lines from \ion{Fe}{} ions to reduce the uncertainty related to the assumption of the elemental composition. The only exception is the \ion{O}{V} \SI{192.91}{\angstrom} line, because it could constrain temperatures below $\log{T} = 5.5$. The event's DEM is higher than that of the background regions at $\log{T} \leq 6.0$, while they are similar for $\log{T} \geq 6.0$. For all three DEMs, we observe a common peak around \SI{2}{\mega\kelvin}, which is probably caused by the line-of-sight integration of the corona, including the large coronal loop standing above \evlbeis. 
 The DEM of the event also shows a second peak around $\log{T}$ = 5.9, due to the abundance of \ion{Fe}{VIII} and \ion{Fe}{IX}. Below $\log{T} = 5.2$ and above  $\log{T} = 6.45$, the DEMs are poorly constrained due to the lack of observed lines. We conclude that \evlbeis barely reaches coronal temperatures, and it is dominated by plasma at TR temperatures.

\subsubsection{Density diagnostics}
\label{sec:results:eis:density}

We derived the electron density of \evlbeis using the density-sensitive \ion{Fe}{IX} 188.49/197.04 line ratio (see Sect. \ref{sec:method:line_ratio}). The ratio is computed with the line intensities of the event, at the effective temperature $\log{T} = 6.0$ of the \ion{Fe}{IX} lines (Table \ref{table:eis_line_radiance}). This high density confirms that we are probably observing a compact, small-scale event, occurring low  in the solar atmosphere.

Figure \ref{fig:results:ratio_fe_9_event} displays the theoretical ratio as a function of the density, as predicted by CHIANTI, with a superposition of the measured one. We obtain a ratio of $0.33\pm 0.04$, which corresponds to a density equal to $n_{\mathrm{e}} =$ \SI[separate-uncertainty = true]{1.8(13)e10}{\per\centi\meter\tothe{3}}. 

For comparison, we inferred the density of the coronal loop in the foreground of the event at  the position occupied by the event. We use the \ion{Fe}{XIII} (203.80 + 203.83)/202.04 line ratio (see Sec. \ref{sec:method:line_ratio}) and assumed a temperature $\log{T} = 6.25$. We obtained a loop density of \SI[separate-uncertainty = true]{4.3(3)e8}{\per\centi\meter\tothe{3}}. This value is typical of coronal loops \citep{Reale2014} and supports our idea that this loop is higher above \evlbeis. 


\section{Discussion}
\label{sec:discussion}

In this work, we have produced evidence that small EUV brightenings detected by \hrieuv in the QS are, for the most part, dominated by plasma at chromospheric and TR temperatures. This is, at least, true for our observed features. As such, their direct contribution to coronal heating is insignificant. 
To obtain these results, we analyzed three QS observation sequences coordinated between multiple instruments. We identified nine events with \hrieuv, AIA and SPICE, and three events with \hrieuv and EIS. We resume our main findings in the following.

Most of the smaller EUV brightenings visible in SPICE (up to \SI{2.5}{\mega\meter}),  for temperatures below $\log{T} = 5.6$, have a higher EM than their background/foreground. The larger event (\evlbeis, \SI{6.2}{\mega\meter}) identified in EIS, barely reaches coronal temperatures, and it is also dominated by plasma at TR temperatures. We estimated the electron density to be equal to $n_{\mathrm{e}} =$ \SI[separate-uncertainty=true]{1.8(13)e10}{\per\centi\meter\tothe{3}}. This value is an order of magnitude above the typical electron densities in the QS near the effective temperatures of the \ion{Fe}{IX} \citep[$n_{\mathrm{e}} \approx 10^9$ \SI{}{\per\centi\meter\tothe{3}} at $T =$ \SI{1.12e6}{\kelvin,}][]{Dere_2020}.

We investigated the light curves from lines at different temperatures (\hrieuv, SPICE, and AIA), and obtained two main results: (1) For most events, and especially for the smaller ones, all of the intensity peaks are co-temporal. 
This is consistent with the results of \cite{Dolliou_2023} using AIA data only. (2), For most events, the \hrieuv light curve follows the behavior of \ion{O}{VI}, including the co-temporality of the peak in intensity (for instance, see Fig.\,\ref{fig:results:all_3_events_light_curves}). Thus, these events are probably dominated by TR temperatures over their lifetime. To conclude, we present evidence that the thermal behavior of these events reflects that of cool features.

The events observed with SPICE showed strong emission in the TR lines, but rarely in the \ion{Ne}{VIII} and never in the \ion{Mg}{IX} lines. These conclusions were also found in the three events analyzed by \cite{Huang_2023}, which barely produced emission in \ion{Ne}{VIII}. While most of the events detected by \hrieuv were clearly identified in SPICE, we can report two exceptions in \seqa. One of them is a small (\SI{0.3}{\mega\meter}) and short-lived one-minute, dot-like event (Fig. \ref{fig:annex:method:event_undetected_spice1}). It is only detected in \hrieuv and possibly in the \ion{C}{iii} line. The other event is a larger (\SI{1.9}{\mega\meter}) loop-like event of five-minute lifetime (Fig.\,\ref{fig:annex:method:event_undetected_spice2}). It is detected in \hrieuv and barely in AIA 171. It is unclear whether the lack of signal can be explained by a low EM, or because the event is dominated by plasma at coronal temperature not covered by the SPICE lines. Although being a minority, these events detected only in \hrieuv should be further investigated, as they may be good candidates for reaching coronal temperatures and possibly participate, even a small amount, to sustaining the corona.

\cite{Nelson_2023} analyzed 15 EUV brightenings on \seqa between 00:04:12 UT and 00:33:33 UT. They used datasets from \hrieuv, AIA, and the IRIS. Our event's dataset in \seqa (Table \ref{table:event_properties}) does not overlap with theirs. IRIS covered chromospheric lines, from $\log{T} = $ 4.0 to 4.9. The authors compared the \hrieuv light curves with those of the \ion{Mg}{II} ($\log{T} = 4.3$) and \ion{Si}{IV} lines ($\log{T} = 4.9$). They found no typical behavior, the light-curve peaks having either positive, negative, or no time delays. Our results showed that the intensity peaks in the SPICE TR light curves were mostly co-temporal with the \hrieuv ones (Fig. \ref{fig:results:all_3_events_light_curves} (a) and (c) for instance).  The \ion{C}{iii} lines in SPICE ($\log{T} = 4.8$) are close to the temperature emissivity of \ion{Si}{IV}, while \ion{H}{I} Lyman-$\beta$ is close to that of \ion{Mg}{II}. The delays between the \hrieuv and the \ion{C}{III} intensity peaks are within \SI{25}{\second} of one another for the majority of events, although we did detect delays up to \SI{300}{\second} for \evh (see Fig.\,\ref{fig:annex080322:E7_E8_E9_lc_spice_aia_hrieuv_normed}). We also observed delays below \SI{1}{\min} between the intensity peaks in \ion{H}{I} Lyman-$\beta$, \ion{C}{III}, and \ion{O}{VI} for \eva (Fig.\,\ref{fig:annex080322:E1_E3_E5_lc_spice_aia_hrieuv_normed}). Positive and negative delays are also observed between the peaks in the \ion{Ne}{VIII} and the \ion{O}{VI} light curve on \evb, \evc, and \evg (Fig. \ref{fig:results:all_3_events_light_curves} (b), \ref{fig:annex080322:E1_E3_E5_lc_spice_aia_hrieuv_normed} (b), \ref{fig:annex080322:E7_E8_E9_lc_spice_aia_hrieuv_normed} (b)). They might be signatures of plasma cooling or heating passing through the temperature sensitivity of the lines. As for \cite{Nelson_2023} and \cite{Dolliou_2023}, we conclude that the population of the events are not uniform and can reach different temperatures. The short delays between the peaks of the TR and the \hrieuv light curves suggest that the lines observed in SPICE can explain what is seen in \hrieuv.

Our results are in line with the finding that the interpretation of what is observed in EUV imagers is not straightforward. It is known that, under certain plasma conditions, the cool emission can dominate the measured signal, even though the peak of the response function of the band is centred at a coronal line. This has been investigated for \hrieuv \citep{Tiwari2022} and for other imagers such as AIA \citep{Schonfeld_2020} or TRACE \citep{Winebarger_2002}.  

The density of \evlbeis was estimated to \SI[separate-uncertainty = true]{1.8(13)e10}{\per\centi\meter\tothe{3}} at $\log{T} = 6.0$, (see Sect. \ref{sec:results:eis:density}). We can compare this value with the upper limits on the density of \evmethod, equal to \SI{2.3e9}{\per\centi\meter\tothe{3}} at $\log{T} = 6.0$ (Sect. \ref{sec:results:spice:loci}).  We conclude that the smaller events detected with SPICE have less plasma at coronal temperature compared with \evlbeis.

The properties of the brightenings are similar to the loops previously observed with Hi-C in the intermoss of an active region  \citep{Winebarger2013}, even though much smaller in length. Their electron temperature ($\log{T} \approx 5.4$) and density ($\log{n} \sim 10$) closely match our results. The authors also observed close to no delay between the peaks of the AIA and the  Hi-C light curves. Furthermore, modellings showed that the heating events are low-lying in the atmosphere, up to \SI{2}{} to \SI{3}{\mega\meter} above the photosphere. These heights are also consistent with EUV brightenings detected with \hrieuv \citep{Zhukov2021}. In addition to active regions, cool (less than \SI{0.5}{\mega\kelvin}) and low-lying (up to \SI{5}{\mega\meter}) loops have also been observed in the QS with IRIS \citep{Hansteen2014}. Magnetic reconnexions involving these types of loops are good candidates to explain the physical origin of the events analysed in our work. They concentrated around the chromospheric network \citep{feldman1999, Almeida2007}, which is consistent with \hrilya observations by \cite{Berghmans2021}. Signatures of magnetic flux cancellations associated with EUV brightenings \citep{Kahil_2022} already suggested that magnetic reconnexion triggered by the footpoints may play an important role. With this work, we argue that the electron temperature (below \SI{1}{\mega\kelvin}) and density (above \SI{5e9}{\per\centi\meter\tothe{3}}) of the EUV brightenings observed with \hrieuv is also consistent with cool loops subjected to impulsive heating.

We showed that the brightenings analyzed in our work barely reach coronal temperatures. While some of them can reach temperatures up \SI{1}{\mega\kelvin} (for instance \evb and \evlbeis), they are, for the most part, dominated by TR plasma. As this study was limited to 12 events, more statistical work is necessary, to verify if these observations can be generalized. If this is the case, this would mean that these events do not directly contribute to the coronal heating, as they do not reach coronal temperatures. Additional modelling work us  necessary to reproduce the properties of the EUV brightenings \citep{Winebarger2013,Hansteen2014}. Such modelling will have to take into account the new spectroscopic constraints for these events, set by the latest results from  the Solar Orbiter observations.


\begin{acknowledgements}
The authors thank the anonymous referee for the useful comments that helped to improve the manuscript. The authors gratefully thank the AIA, the EIS, the EUI, and the SPICE teams for the planning of the observations. The authors thank F. Auchère for his help in the  preparation of SPICE data. A.D. acknowledges funding from CNES and the Ecole Doctorale Ondes et Matière (EDOM). Solar Orbiter is a space mission of international collaboration between ESA and NASA, operated by ESA. The EUI instrument was built by CSL, IAS, MPS, MSSL/UCL, PMOD/WRC, ROB, LCF/IO with funding from the Belgian Federal Science Policy Office (BELSPO/PRODEX PEA C4000134088); the Centre National d’Etudes Spatiales (CNES); the UK Space Agency (UKSA); the Bundesministerium für Wirtschaft und Energie (BMWi) through the Deutsches Zentrum für Luft- und Raumfahrt (DLR); and the Swiss Space Office (SSO). The development of SPICE has been funded by ESA member states and ESA. It was built and is operated by a multi‑national consortium of research institutes supported by their respective funding agencies: STFC RAL (UKSA, hardware lead), IAS (CNES, operations lead), GSFC (NASA), MPS (DLR), PMOD/WRC (Swiss Space Office), SwRI (NASA), UiO (Norwegian Space Agency).This work used data provided by the MEDOC data and operations centre (CNES / CNRS / Univ. Paris-Saclay. Hinode is a Japanese mission developed and launched by ISAS/JAXA, collaborating with NAOJ as a domestic partner, NASA and UKSA as international partners. Scientific operation of the Hinode mission is conducted by the Hinode science team organized at ISAS/JAXA. This team mainly consists of scientists from institutes in the partner countries. Support for the post-launch operation is provided by JAXA and NAOJ (Japan), UKSA (U.K.), NASA, ESA, and NSC (Norway). This research was supported by the International Space Science Institute (ISSI) in Bern, through ISSI International Team project \#23-586 (Novel Insights Into Bursts, Bombs, and Brightenings in the Solar Atmosphere from Solar Orbiter). This research used version 0.7.4 \citep{barnes_2021_5606094} of the aiapy open source software package \citep{Barnes2020}. This research used version 5.0.1 (\citep{stuart_j_mumford_2023_8384174} of the SunPy open source software package \citep{sunpy_community2020}. This research used the open source sospice Python package. This research used CHIANTI version 10.1. CHIANTI is a collaborative project involving George Mason University, the University of Michigan (USA), University of Cambridge (UK) and NASA Goddard Space Flight Center (USA). 

\end{acknowledgements}

\bibliographystyle{aa}
\bibliography{Biblio.bib}

\begin{appendix}
\section{Alignment algorithm}
\label{sec:annex:alignment}
\begin{figure*}[h!]
    \centering
    \includegraphics[width=\textwidth]{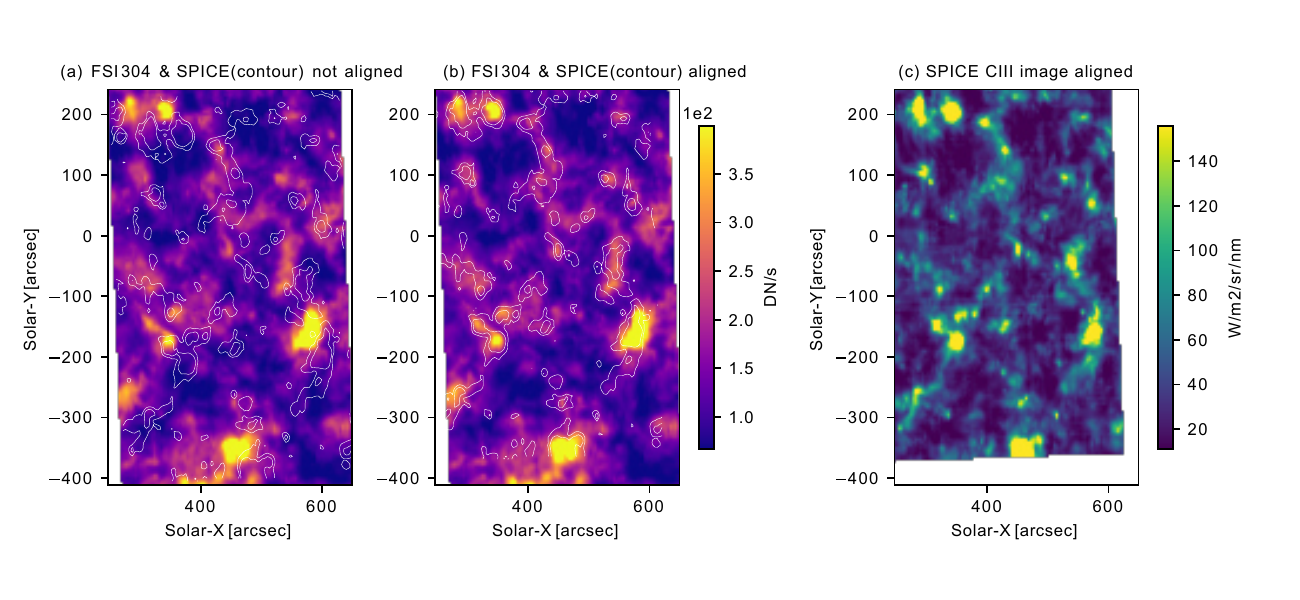}
    \caption{Example of the result of the alignment between SPICE and FSI for \seqb. (a) the FSI 304 synthetic raster is displayed in the background along with the 80 and 90\,\% percentile as contours of the not aligned \ion{C}{III} SPICE map. (b) The same composite image after the alignment. (c) SPICE \ion{C}{III} map corrected after the alignment.}
    \label{fig:annex:alignment}
\end{figure*}

We developed tools to co-register the SPICE and the \hrieuv sequences. For \seqa, the SPICE \ion{Ne}{VIII} image of the context raster is co-aligned with an \hrieuv synthetic raster. The latter is built by selecting the \hrieuv image closest in time to each of the raster steps. We build SPICE images by summing the intensity over the spectral window. The same technique is applied to \seqb with the \ion{C}{III} line and the FSI 304 sequence. 

Figure \ref{fig:annex:alignment}, as an example, illustrates  the result for the \seqb data. 
We co-align the SPICE context raster with the FSI 304 synthetic raster, using a cross-correlation method. The offset found with this correlation is then applied to the values of the pointing information in the FITS headers.

The \hrieuv temporal sequences are co-registered with FSI images. For \seqa and \seqc, we co-align the \hrieuv images closest in time to each of the \fsieuv images in the dataset. The cross-correlation technique is the same as the one discussed for SPICE. We then apply a correction to the spacecraft jitter, inspired by \cite{Chitta_2022}. The \hrieuv dataset is divided into sub lists, with the first images being those already co-aligned with \fsieuv images. Every \hrieuv image is then co-aligned with the first image of its sub-list. As a last step, we corrected remains of the jitter by again cross-correlating the \hrieuv sequence within sub-lists, this time with overlapping images between the sub-lists. 

For \seqb (2022 March 17, 00:18\,--\,00:48 UTC), only FSI 304 images at one-minute cadence were available before 00:16 UT, and we had to adapt the co-registration procedure. We first co-align an \hrieuv and an \fsieuv image at 09:50:04 UTC, to locate the position of the \hrieuv field of view on that of FSI. As \fsieuv and FSI 304 share the same field of view, we can co-align the \hrieuv images with the FSI 304 image closest in time. 

These methods are available through the \software{\mbox{euispice\_coreg}}\footnote{\url{https://github.com/adolliou/euispice_coreg}} Python package.

\section{Additional events in \seqa}
\label{annex:080322_events}

\begin{figure*}
    \centering
    \includegraphics[width=\textwidth]{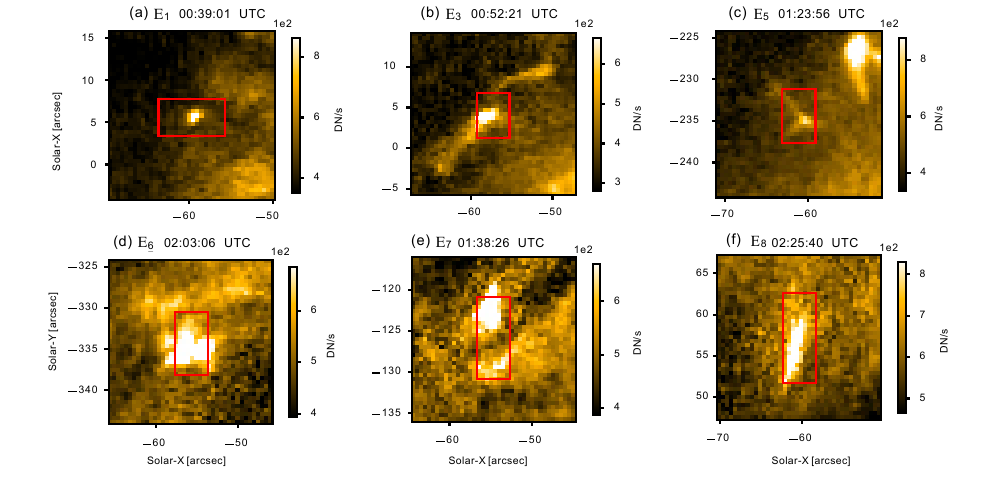}
    \caption{\hrieuv images centred around the events detected in \seqa and \seqb (see Table \ref{table:event_properties}), that are not discussed in  Sect. \ref{sec:results}. The red rectangles are the event regions, that are defined in section \ref{sec:method:event_detection}.}
    \label{fig:annex:all_08_events_hrieuv_imshow}
\end{figure*}

\begin{figure*}
    \centering   \includegraphics{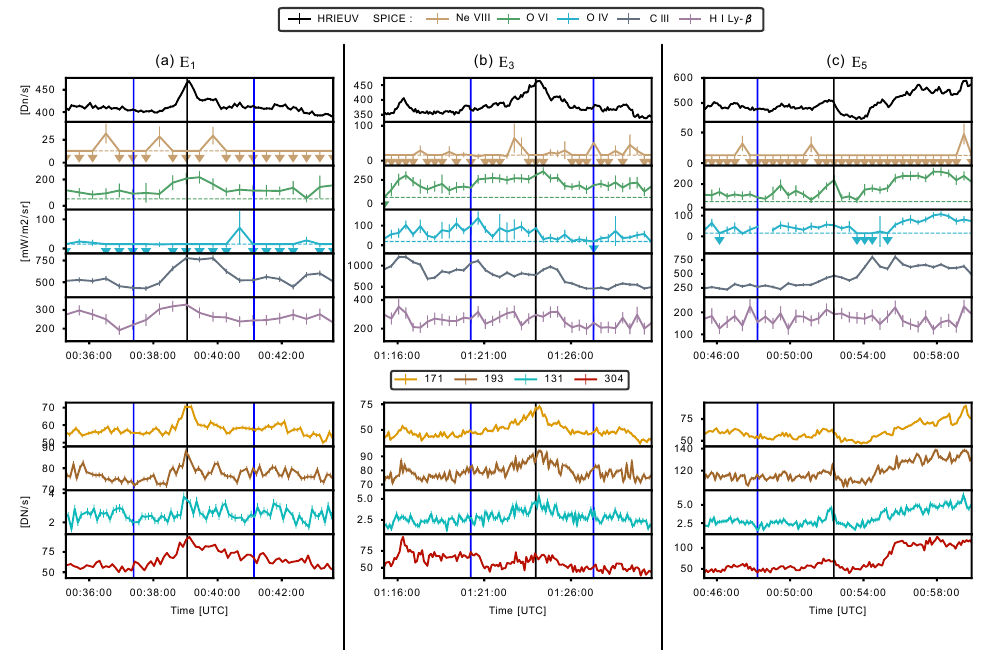}
    \caption{\hrieuv, SPICE, and AIA light curves for \eva (a), \evc (b) et \eve (c). The vertical black lines indicate the SPICE times closest to the \hrieuv peaks, when the EMs are estimated in Fig. \ref{fig:annex080322:E1_E3_E5_loci_article_1.00E+09}. The two background time intervals of two minutes, defined in Sect. \ref{sec:method:emloci_dem}, are represented by two vertical blue lines in each panel. The left blue line indicates the end of B$_1$ (before the event), and the right blue line indicates the start of B$_2$ (after the event). \eve only includes $B_1$, as it is before another larger event. The upper limits of the SPICE radiance are shown as horizontal dotted lines. When the line is not detected above noise levels, the radiance is being replaced by its upper limit, and the point is marked by a downward pointing arrow. This figure is similar to Fig. \ref{fig:results:all_3_events_light_curves} }
\label{fig:annex080322:E1_E3_E5_lc_spice_aia_hrieuv_normed}
\end{figure*}

\begin{figure*}
    \centering   \includegraphics{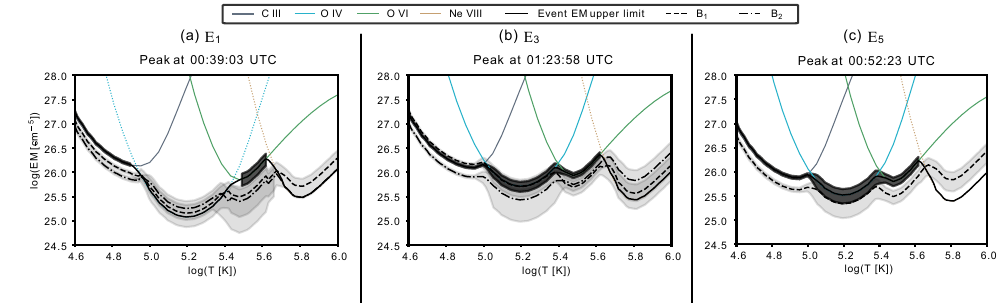}
    \caption{ EM estimated with the LOCI method of \eva (a), \evc (b), and \evd (d) at the time of their \hrieuv peaks, using four SPICE lines. The EMs of the event are shown for each SPICE line with colored curves. Dotted curves indicate that the radiance is an upper limit. The black full curves display the EM for the event. The dotted and the dashed black lines are the EMs for the B$_1$ and B$_2$ background time intervals. B$_1$ also includes upper limits in \ion{O}{IV} for \evb and in \ion{Ne}{VIII} for \evd.  The black (event) and grey (B$_1$ and B$_2$) regions delimit the 2$\sigma$ uncertainty of the EMs. This Figure is similar to Fig. \ref{fig:results:all_3_events_loci}. }
\label{fig:annex080322:E1_E3_E5_loci_article_1.00E+09}
\end{figure*}

\begin{figure*}
    \centering
\includegraphics{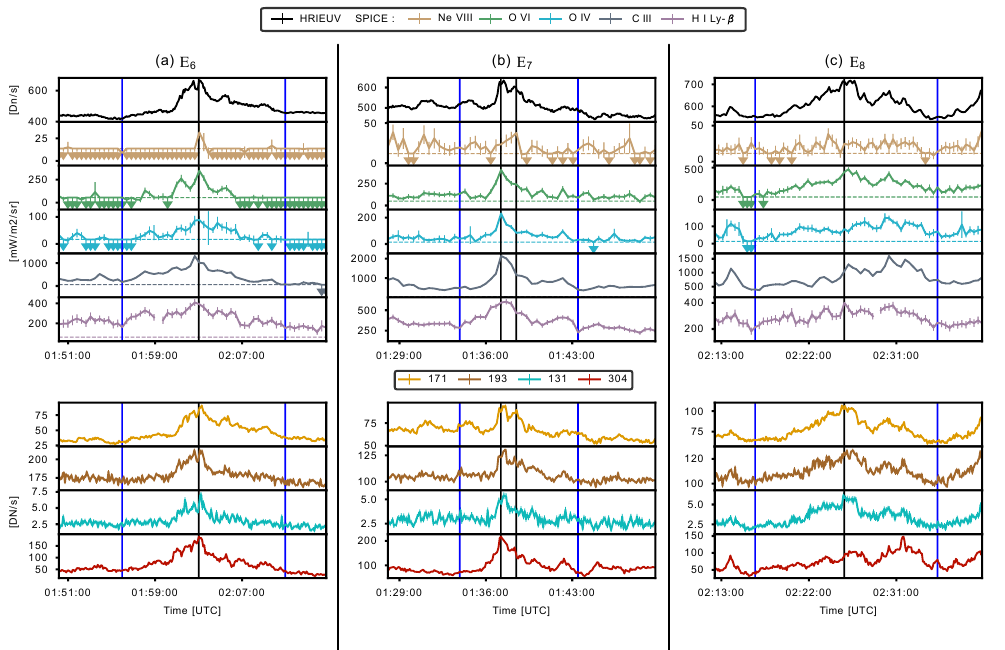}
    \caption{Same as Fig. \,\ref{fig:annex080322:E1_E3_E5_lc_spice_aia_hrieuv_normed}, for \evf, \evg, and \evh.}\label{fig:annex080322:E7_E8_E9_lc_spice_aia_hrieuv_normed}
\end{figure*}

\begin{figure*}
    \centering
\includegraphics{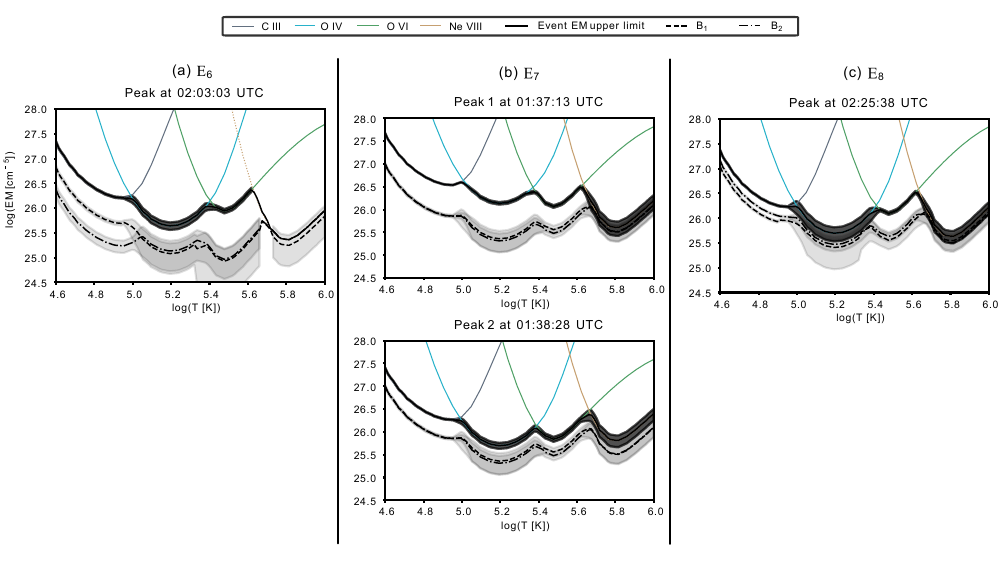}
    \caption{ Same as Fig. \ref{fig:annex080322:E1_E3_E5_loci_article_1.00E+09}, for \evf, \evg, and \evh.}\label{fig:annex080322:E6_E7_E8_loci_article_1.00E+09}
\end{figure*}

\begin{figure*}
    \includegraphics[width=\textwidth]{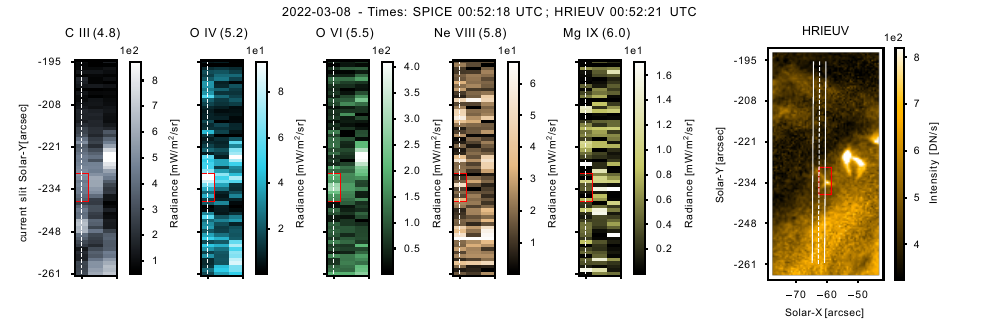}
    \caption{
    The \hrieuv (left panel) and the SPICE images are centred around \eve, at the time of its peak in \hrieuv intensity on \seqb. The position of the slit is displayed as a dashed line on the \hrieuv and the SPICE images. The two white lines in \hrieuv delimit the field of view of the SPICE slit. The red rectangle is the event region selected, and it is defined in Sect. \ref{sec:method:identification_SPICE}. The  temperature ($\log{T}$) of the maximum emissivity of each SPICE line is indicated within parentheses above each image. The latitude on the $y$-axis of the SPICE images refers to the position of the slit marked with a white dashed line. This Figure is similar to Fig. \ref{fig:method:bright_dot_170322_all_event_spice_imshow_peak2}.}
\label{fig:annex:method:E5_small_dot_06_spice_imshow_00064}
\end{figure*}

In this Appendix, we present the results for the events in \seqa (Table \ref{table:event_properties}) that were not discussed in Sect. \ref{sec:results:spice}. Figure \ref{fig:annex:all_08_events_hrieuv_imshow} shows \hrieuv images centred around the events at the time of one of their \hrieuv peak. The results of their light curves and the EM LOCI method are displayed for \eva, \evc, \eve (Fig.\,\ref{fig:annex080322:E1_E3_E5_lc_spice_aia_hrieuv_normed} and \ref{fig:annex080322:E1_E3_E5_loci_article_1.00E+09}), and for \evf, \evg, \evh (Fig. \ref{fig:annex080322:E7_E8_E9_lc_spice_aia_hrieuv_normed} and \ref{fig:annex080322:E6_E7_E8_loci_article_1.00E+09}). As an example, \eve seen with \hrieuv and SPICE is displayed in Figure \ref{fig:annex:method:E5_small_dot_06_spice_imshow_00064}, and as a movie in the multimedia materials.

For most events, the \ion{O}{vi} intensity peaks are co-temporal with the those of \hrieuv. Exceptions include \evg (Fig.\,\ref{fig:annex080322:E7_E8_E9_lc_spice_aia_hrieuv_normed} (b)), that has two \hrieuv peaks co-temporal with either \ion{O}{vi} or \ion{Ne}{viii}. Similar results were found for \evb in Figure \ref{fig:results:all_3_events_light_curves} (c), and discussed in section \ref{sec:results:spice:lc}. We suggest that these two peaks in \hrieuv are caused by either the TR or the coronal contribution to the response function of the channel. Delays between peaks in \hrieuv and in \ion{C}{iii} are also observed. They can reach up to \SI{300}{\second} for \evh, which is the largest event detected in \seqa (Fig.\,\ref{fig:annex080322:E7_E8_E9_lc_spice_aia_hrieuv_normed} (c)). In that case, the \ion{C}{iii} intensity peak is after those of \hrieuv, suggesting a  plasma cooling to chromospheric temperature.

The EMs of all of the events (Fig.\,\ref{fig:annex080322:E1_E3_E5_loci_article_1.00E+09} and \ref{fig:annex080322:E6_E7_E8_loci_article_1.00E+09}) shows that they are dominated by plasma below $\log{T} = 5.6$. Only the EM of \evg is clearly enhanced for $\log{T} > 5.6$ (Fig.\,\ref{fig:annex080322:E6_E7_E8_loci_article_1.00E+09} (b), bottom panel). We conclude the events are dominated by plasma at chromospheric and TR temperatures, with some of them barely reaching coronal temperatures.

\newpage

\newpage

\begin{figure*}
    \includegraphics[width=\textwidth]{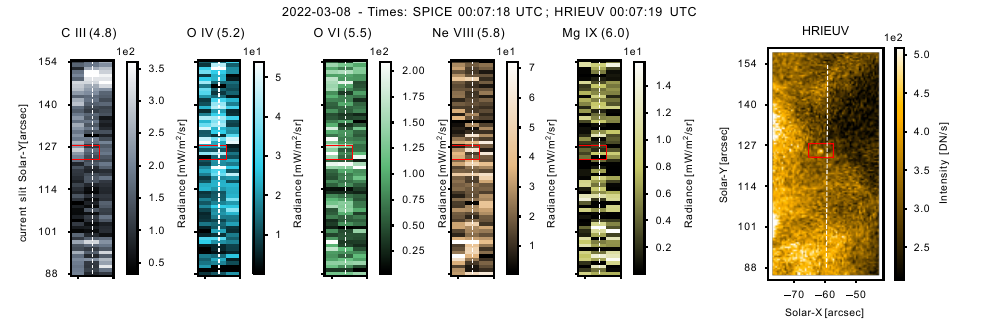}
    \caption{Similar to Fig.\,\ref{fig:annex:method:E5_small_dot_06_spice_imshow_00064}, for a small (\SI{0.3}{\mega\meter}) dot-like event not clearly identified in SPICE.  }
\label{fig:annex:method:event_undetected_spice1}
\end{figure*}

\begin{figure*}
    \includegraphics[width=\textwidth]{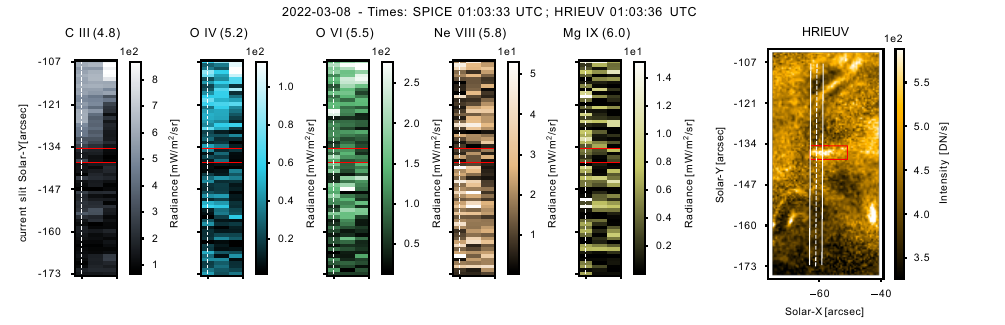}
    \caption{Similar to Fig.\,\ref{fig:annex:method:E5_small_dot_06_spice_imshow_00064}, for a loop-like event (\SI{1.9}{\mega\meter}) not clearly identified in SPICE.}
\label{fig:annex:method:event_undetected_spice2}
\end{figure*}

\newpage
\section{Supplementary Material}

\begin{table*}
\centering          
\begin{tabular}{l || c c c || c c c || c c c } 

Line & \multicolumn{3}{c||}{Event region} & \multicolumn{3}{c||}{Region 1} & \multicolumn{3}{c}{Region 2} \\
\hline
   & $I_{\mathrm{ev}}$ & $I_{\mathrm{th}}/I_{\mathrm{ev}}$ & $T_{\mathrm{eff}}$ & $I_{\mathrm{b1}}$ & $I_{\mathrm{th}}/I_{\mathrm{b1}}$ & $T_{\mathrm{eff}}$ & $I_{\mathrm{b2}}$ & $I_{\mathrm{th}}/I_{\mathrm{b2}}$ & $T_{\mathrm{eff}}$ \\
 \hline 
 \ion{O}{V$^{\star}$} 192.91 & $29.75 \pm 1.97$ & $0.95$ & $5.40$ & $14.41\pm 2.98$ & $0.90$ & $5.40$ & $16.86 \pm 1.93$ & $0.71$ & $5.40$ \\
\ion{Fe}{VII} 195.39 & $12.62 \pm 1.15$ & $0.74$ & $5.52$ & $6.60\pm 1.08$ & $0.69$ & $5.52$ & $6.44 \pm 0.93$ & $0.65$ & $5.52$ \\
\ion{Fe}{VIII$^{\star}$} 194.66 & $14.41 \pm 1.03$ & $1.08$ & $5.81$ & $6.56\pm 0.80$ & $1.09$ & $5.781$ & $8.41 \pm 0.86$ & $0.86$ & $5.82$ \\
\ion{Fe}{VIII} 186.60 & $51.97 \pm 3.80$ & $0.91$ & $5.82$ & $26.41\pm 2.36$ & $0.83$ & $5.79$ & $20.58 \pm 2.72$ & $1.08$ & $5.83$ \\
\ion{Fe}{VIII} 185.21 & $66.17 \pm 4.17$ & $1.04$ & $5.83$ & $28.66\pm 2.54$ & $1.10$ & $5.80$ & $31.21 \pm 3.00$ & $1.03$ & $5.83$ \\
\ion{Fe}{IX}$^{\star}$ 188.49 & $32.47 \pm 1.53$ & $0.88$ & $6.00$ & $11.44\pm 1.07$ & $1.00$ & $6.02$ & $17.00 \pm 1.19$ & $0.93$ & $6.06$ \\
\ion{Fe}{IX} 197.86 & $10.73 \pm 1.04$ & $1.25$ & $6.01$ & $5.83\pm 0.82$ & $0.94$ & $6.03$ & $5.60 \pm 0.79$ & $1.39$ & $6.07$ \\
\ion{Fe}{X} 184.54 & $78.96 \pm 4.54$ & $1.00$ & $6.09$ & $37.33\pm 3.57$ & $1.08$ & $6.14$ & $71.11 \pm 4.49$ & $0.93$ & $6.13$ \\
\ion{Fe}{X}$^{\star}$ 190.04 & $27.75 \pm 1.47$ & $0.93$ & $6.09$ & $15.22\pm 1.17$ & $0.87$ & $6.14$ & $30.78 \pm 1.90$ & $0.70$ & $6.13$ \\
\ion{Fe}{XI}$^{\star}$ 188.21 & $248.20 \pm 6.65$ & $1.04$ & $6.17$ & $183.67\pm 5.99$ & $1.04$ & $6.21$ & $255.50 \pm 6.49$ & $1.05$ & $6.18$ \\
\ion{Fe}{XII}$^{\star}$ 195.12 & $493.54 \pm 5.71$ & $0.99$ & $6.22$ & $456.49\pm 5.70$ & $0.99$ & $6.24$ & $527.55 \pm 6.22$ & $0.98$ & $6.22$ \\
\ion{Fe}{XIII}$^{\star}$ 202.12 & $474.31 \pm 8.87$ & N/A & $6.25$ & $502.77\pm 8.93$ & N/A & $6.26$ & $520.98 \pm 9.35$ & N/A & $6.25$ \\
\ion{Fe}{XIII}$^{\star}$ 203.83 & $229.28 \pm 10.39$ &N/A & $6.25$ & $207.45\pm 9.29$ & N/A & $6.26$ & $254.79 \pm 10.74$ & N/A & $6.25$ \\
\ion{Fe}{XIV} 264.79 & $209.35 \pm 8.24$ & $0.98$ & $6.29$ & $223.93\pm 8.20$ & $0.93$ & $6.29$ & $256.47 \pm 8.99$ & $0.82$ & $6.29$ \\
\ion{Fe}{XV} 284.16 & $1053.09 \pm 24.83$ & $1.01$ & $6.35$ & $980.05\pm 20.96$ & $1.02$ & $6.33$ & $1018.45 \pm 25.12$ & $1.05$ & $6.34$ \\
\ion{Fe}{XVI} 262.98 & $40.01 \pm 3.31$ & $0.90$ & $6.41$ & $35.35\pm 3.00$ & $0.87$ & $6.41$ & $52.83 \pm 3.80$ & $0.67$ & $6.41$ \\
\hline   
\end{tabular}
\vspace{0.5cm}
\caption{EIS line radiances and uncertainties (in \SI{}{\milli\watt\per\second\square\per\meter\per\steradian}), fitted from the spectra averaged over the event \evlbeis regions (ev), and the background 1 (b1) and two (b2) regions. These values are used for the calculation of the DEM shown in Fig.\,\ref{fig:results:dem_eis_coronal_no_o5} (b).  For each region, the ratio of the reconstructed over the observed radiance is given for each line, along with the logarithm of their effective temperature in Kelvin. The wavelengths are in \SI{}{\angstrom}, and the lines marked with a star are blended. The \ion{Fe}{XIII} lines are not used for the DEM inversion, so their reconstructed to observed ratio is marked  "N/A". }         
\label{table:eis_line_radiance}
\end{table*}

\begin{figure*}
    \includegraphics[width=\textwidth]{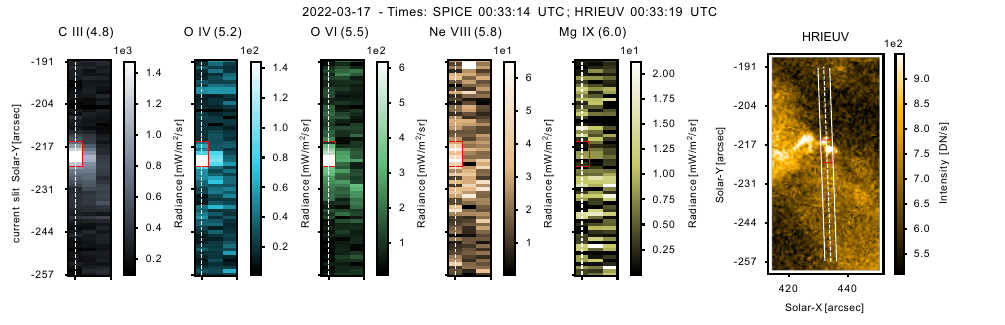}
    \caption{Same as Fig\,\ref{fig:method:bright_dot_170322_all_event_aia_imshow_peak2}, for the \hrieuv intensity peak of \evmethod at 00:33:19 UTC.}
\label{fig:annex:170322_bright_dot_02_peak1_00146}
\end{figure*}

\begin{figure*}
    \centering
    \includegraphics[width=\textwidth]{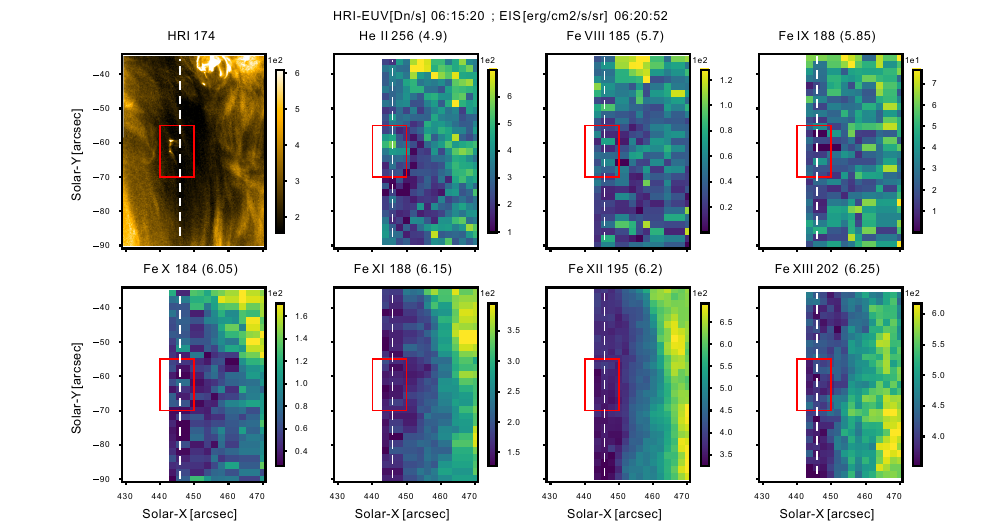}
    \label{fig:annex:small_loop_3_imshow}
    \caption{Images of \hrieuv and EIS centred around \evm, visible in the \ion{He}{II} \SI{256}{\angstrom} line. The white dotted lines show the position of the slit at 06:34:41 UTC, and the red rectangle is an area centred around \evm. The temperatures ($\log{T}$) of the peak emissivity are indicated within parentheses for each line.}
\end{figure*}

\label{sec:annex:spice_fitting}
\begin{figure*}
    \centering
    \includegraphics{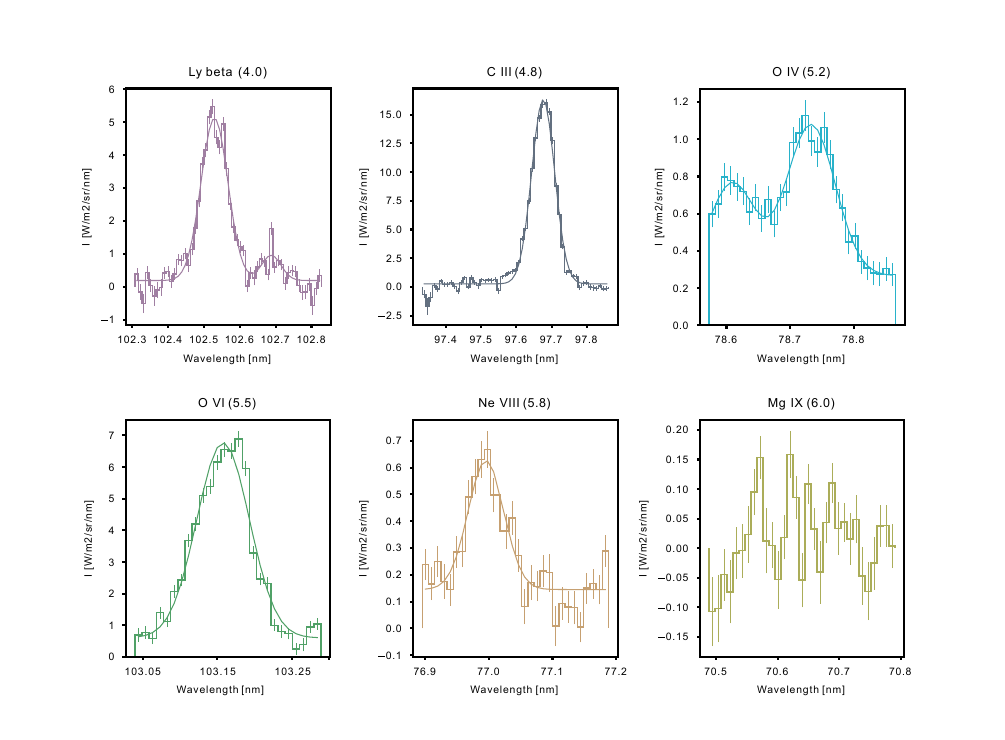}
    \caption{Averaged spectra over the event mask of \evmethod at 00:36:09 UTC (Fig.\,\ref{fig:method:bright_dot_170322_all_event_spice_imshow_peak2}), along with the fitting of Gaussian functions. No \ion{Mg}{ix} line is detected. The temperature ($\log{T}$) of the peak emissivity is indicated within parentheses for each line.}
    \label{fig:annex:time_peak_1}
\end{figure*}

\begin{figure*}
    \includegraphics[width=0.49\textwidth]{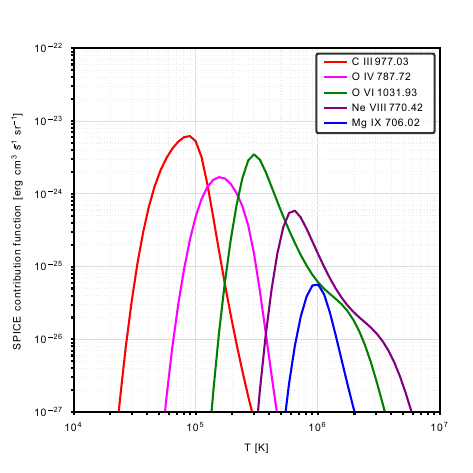}
    \includegraphics[width=0.49\textwidth]{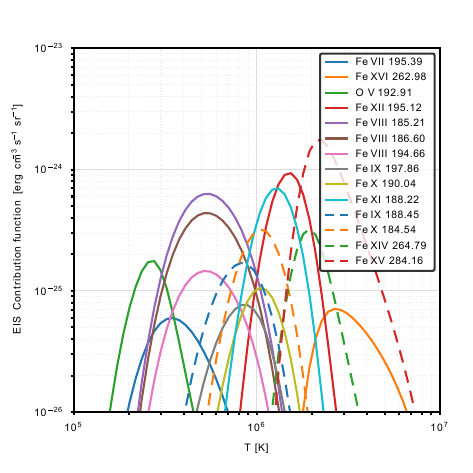}
    \caption{Contribution function of the SPICE (left) and EIS (right) lines computed with CHIANTI V10.1, assuming a coronal abundance \citep{Asplund_2021}, the recommended ionisation fraction by CHIANTI, and an electron density equal to $n_{\mathrm{e}} = $ \SI{1e9}{\per\centi\meter\tothe{3}}. }
    \label{fig:annex:goffnt}
\end{figure*}

\end{appendix}
\end{document}